\documentclass[aps,prx,twocolumn,superscriptaddress,english]{revtex4-1}
\usepackage{amssymb}
\usepackage{graphicx}
\usepackage{amsmath}
\usepackage{amsthm}
\usepackage{amsfonts}
\usepackage{txfontsb}
\usepackage{color}
\usepackage{bm}
\usepackage{bbm}
\usepackage{url}
\usepackage[driverfallback=dvipdfm]{hyperref}
\hypersetup{pdfpagemode=FullScreen,colorlinks=true,breaklinks,urlcolor=blue,linkcolor=blue,citecolor=blue}

\makeatletter
\def\iddots{\mathinner{\mkern1mu\raise\p@
    \hbox{.}\mkern2mu\raise4\p@\hbox{.}\mkern2mu
    \raise7\p@\vbox{\kern7\p@\hbox{.}}\mkern1mu}}
\def\adots{\mathinner{\mkern2mu\raise\p@\hbox{.} 
 \mkern2mu\raise4\p@\hbox{.}\mkern1mu
 \raise7\p@\vbox{\kern7\p@\hbox{.}}\mkern1mu}}
\makeatother

\begin{document}

\global\long\def\id{\mathbbm{1}}
\global\long\def\ui{\mathbbm{i}}
\global\long\def\ud{\mathrm{d}}


\title{Unconventional Floquet topological phases from quantum engineering of band inversion surfaces}

\author{Long Zhang}
\email{lzhangphys@hust.edu.cn}
\affiliation{School of Physics and Institute for Quantum Science and Engineering, Huazhong University of Science and Technology, Wuhan 430074, China}
\affiliation{International Center for Quantum Materials, School of Physics, Peking University, Beijing 100871, China}
\affiliation{Collaborative Innovation Center of Quantum Matter, Beijing 100871, China}
\author{Xiong-Jun Liu}
\email{xiongjunliu@pku.edu.cn}
\affiliation{International Center for Quantum Materials, School of Physics, Peking University, Beijing 100871, China}
\affiliation{Collaborative Innovation Center of Quantum Matter, Beijing 100871, China}
\affiliation{International Quantum Academy, Shenzhen 518048, China}

\begin{abstract}
Floquet engineering provides a toolbox for the realization of novel quantum phases without static counterparts, while conventionally the realization may rely on the manipulation of complex temporal evolution.
Here we propose a systematic and high-precision scheme to realize unconventional Floquet topological phases
by engineering local band structures in particular momentum subspace called band inversion surfaces (BISs).
This scheme is based on a new bulk-boundary correspondence that for a class of generic $d$-dimensional periodically driven systems,
the local topological structure formed in each BIS uniquely determines the features of gapless boundary modes. By engineering the BIS configuration we demonstrate a highly efficient approach to realize, manipulate, and detect novel Floquet topological phases. 
In particular, we predict a two-dimensional (2D) anomalous Floquet valley-Hall phase which carries trivial global bulk topological invariants but features protected counter-propagating edge states in each quasienergy gap.
The unconventional nature of this novel 2D phase is further illustrated by the examination of edge geometry dependence and its robustness to disorder scattering.
Anomalous chiral topological phases with valley protection in higher dimension are also predicted and studied.
Our systematic and highly feasible scheme opens a new route to realize and engineer unconventional Floquet topological phases for ultracold atoms and other quantum simulators.
\end{abstract}

\maketitle

\section{Introduction}

Exploring various topological phases of quantum matter has been a great driving force for the development of quantum materials and quantum simulators.
In the discovery of topological insulators and topological superconductors,
the bulk-boundary correspondence,
which connects the nontrivial bulk topology to the gapless boundary modes,
has played a fundamental role~\cite{TI_review1,TI_review2,Schnyder2008,Kitaev2009,Moore2010,Sato2017} and enabled the identification of the topological matters. The two-dimensional (2D) and 3D topological insulators~\cite{Konig2007,Hsieh2008,Xia2009},
quantum anomalous Hall phase~\cite{Chang2013}, and topological semimetals~\cite{Xu2015,Lv2015} have been discovered.

Recently, Floquet topological phases in periodically driven systems
have attracted extensive interests~\cite{Cayssol2013,Eckardt2017,Harper2020,Rudner2020,Weitenberg2021,
Lindner2011,Kitagawa2010,Rudner2013,Goldman2014,Nathan2015,Gomez-Leon2013,Lababidi2014,Carpentier2015,Fulga2016,Quelle2017,Roy2017,Yao2017,Zhou2018,Peng2019,Nag2019,Hockendorf2019,Liu2019,Umer2020,Hu2020a,Huang2020,Zhanglong2020}
and been explored in synthetic quantum systems~\cite{Rechtsman2013,Jotzu2014,Flaschner2016,Peng2016,Mukherjee2017,Maczewsky2017,Wintersperger2020}.
A periodically driven system can be characterized by the stroboscopic time evolution operator over one driving period,
whose eigenvalues define a set of quasienergy bands, called the Floquet bands~\cite{Eckardt2017,Harper2020,Rudner2020}.
The periodicity of quasienergy brings birth to phases without static counterparts---the anomalous Floquet topological phases in which
boundary modes appear in 
quasienergy gaps 
but have no direct correspondence to the bulk topology of Floquet bands~\cite{Rudner2013}.
The characterization of a $d$-dimensional ($d$D) anomalous Floquet topological phase can be achieved by topological invariants defined in the higher ($d+1$)D momentum-time space~\cite{Kitagawa2010,Rudner2013,Nathan2015}.
However, this characterization is physically not intuitive and unfavorable for the Floquet engineering of new topological phases. Based on such characterization it is not easy to precisely verify anomalous Floquet topological states in experiment, since the anomalous boundary modes and bulk topology defined in momentum-time space are usually not straightforward to measure.

It was shown recently that while an anomalous Floquet topological phase cannot be interpreted from the $d$D Floquet bands, it can be fully characterized by lower-dimensional topology defined in ($d-1$)D momentum subspaces called band inversion surfaces (BISs)~\cite{Zhanglong2020}. The theory based on BISs has important advantages. First, it does not involve the complex time evolution and provides a minimal characterization scheme for the anomalous Floquet topological phases. Second, this characterization enables a direct probe of the bulk Floquet topology by measuring the physics on lower-dimensional BISs through, e.g., quench dynamics~\cite{Zhanglong2020}.
On the other hand, each BIS represents a local subspace of the Brillouin zone, but the BIS configuration has been so far applied as a whole to characterize the global topology by summing over the contributions from {\it all} BISs~\cite{Zhanglin2018,Zhanglong2019a,YuJiZhang2021,Zhanglong2020}. 
It remains unexplored whether {\it each} lower-dimensional local topological structure has unique correspondence to boundary physics and thus plays a more fundamental role in classifying the topological phases, and
whether the BIS characterization can facilitate the realization of novel Floquet topological phases beyond the conventional characterization theory.

In this article, we investigate the fundamental correspondence between BISs and boundary topological physics, and further propose a systematic and high-precision scheme to realize, manipulate, and detect unconventional Floquet topological phases based on BIS engineering.
By employing the BIS characterization and a dimension reduction approach, we show a new bulk-boundary correspondence
that 
the local topological structure formed in each BIS uniquely determines the full features including the number, chirality, and spin-momentum locked texture of the boundary modes.
The BIS-boundary correspondence reveals the fact that BISs serve as a fundamental element in constructing and characterizing Floquet topological phases. On one hand, it enables a highly feasible scheme to realize and manipulate novel topological states by engineering local band structures; on the other hand, the measurement of ($d-1$)D anomalous boundary modes in such topological phases can be precisely mapped to the detection of ($d-1$)D local BIS configurations.
In particular, we propose the realization and detection of a 2D anomalous valley-Hall phase, which resembles but is different from its static counterpart, and also a 3D anomalous chiral phase. These phases feature protected boundary states but cannot be classified by conventional global topological invariants. Instead, they can be precisely characterized from the local topological structures formed in each BIS, and the underlying exotic topological physics is studied.
These novel phases are experimentally accessible in quantum simulators such as ultracold atoms and solid-state spin systems.

Compared to previous works~\cite{Zhanglin2018,Zhanglong2019a,YuJiZhang2021,Zhanglong2020},
The present work has two important advances that should be noted:
(i) This work provides a more fundamental characterization of Floquet topological phases based on BISs.
All previous works treat BISs as a whole such that only the global topological invariant defined on all BISs is utilized to characterizes the global band topology of static~\cite{Zhanglin2018,Zhanglong2019a,YuJiZhang2021} or Floquet~\cite{Zhanglong2020} systems. By contrast, in the present work we show that the local topological structure formed in each {\it single} BIS also has nontrivial topological consequence; the topological invariant defined on each BIS can uniquely correspond to a gapless edge mode, rendering the novel BIS-boundary correspondence.
(ii) With the BIS-boundary correspondence, we propose systematic and feasible schemes to realize unconventional topological phases which are previously undiscovered and beyond the conventional classification. For example, the new 2D and 3D anomalous Floquet topological phase can be readily achieved based on BIS engineering, and these experimental schemes can be naturally extended to even higher dimensions.

This paper is organized as follows. In Sec.~\ref{sec2}, we briefly review the BIS characterization theory for a generic $d$D periodically driven model.
In Sec.~\ref{sec3}, we derive the key results of BIS-boundary correspondence for both static and Floquet systems.
In Sec.~\ref{sec4}, we propose the systematic scheme to realize and manipulate unconventional Floquet topological phases through BIS engineering, and illustrate it with two concrete examples.
We predict a 2D anomalous valley-Hall phase and a 3D anomalous chiral phase, which are both beyond the conventional characterization but can be precisely characterized and thus detected by topological structures in sub-dimensional BISs.
For the 2D phase, its edge-geometry dependence and disorder robustness are analyzed in detail.
In Sec.~\ref{sec5}, we summarize the conclusion with a discussion.
Some technical details are given in appendices.

\section{Model and Band inversion surfaces}~\label{sec2}

We consider a class of $d$-dimensional ($d$D) periodically driven systems described by the Hamiltonian
\begin{equation}\label{FHam0}
H({\bm k},t)={H}_{s}({\bm k})+V({\bm k},t)\gamma_0,\quad H_{s}({\bm k})=\sum_{i=0}^d h_i({\bm k})\gamma_i
\end{equation}
Here $V(t)=V(t+T)$, and the $\gamma$ matrices satisfying $\{\gamma_i,\gamma_j\}= 2\delta_{ij}{\id}$ are
of dimensionality $n_d = 2^{d/2}$ (or $2^{(d+1)/2}$) if
$d$ is even (or odd).
In odd dimensions, the topological phase necessitates a chiral symmetry protection, which is satisfied by the constraint $V(t_{\rm ref}+t)=V(t_{\rm ref}-t)$ ($0\leqslant t_{\rm ref}<T$)~\cite{Zhanglong2020}.
The total Hamiltonian in Eq.~\eqref{FHam0} characterizes the applying of periodic driving on top of a static $d$D band structure.
In the next section, we shall first reexamine the bulk-boundary correspondence for the static system, and then generalize the results to Floquet topological phases.

Here we briefly review the topological characterization theory based on the concept of band inversion surfaces (BISs)~\cite{Zhanglin2018,Zhanglong2019a,YuJiZhang2021,Zhanglong2020}.
For the static Hamiltonian $H_s({\bm k})$,
we take $h_{0}({\bm k})$ to describe the band dispersion, and denote the remaining components as a spin-orbit (SO) field ${\bm h}_{\mathrm{so}}({\bm k})\equiv(h_{1},\dots,h_{d})$, which depicts the coupling between the $n_d$ bands.
Without the SO coupling, energy bands are inverted and cross on $(d-1)$D BISs, defined by
\begin{equation}
\mathrm{BIS}\equiv\{{\bm k}\vert h_{0}({\bm k})=0\}.
\end{equation}
The SO field opens the gap at BISs.
Topological charges reside at the nodes where ${\bm h}_{\mathrm{so}}({\bm k}^c)=0$, with charge value given by
the winding of ${\bm h}_{\mathrm{so}}({\bm k})$ around ${\bm k}={\bm k}^c$.
The topology of the bands below the gap is characterized by the total charges enclosed by BISs, namely the charges in the region where $h_0<0$~\cite{Zhanglong2019a}.
For the case that ${\bm h}_{\mathrm{so}}({\bf k})$ is linear when approaching the node, the charge is
characterized by a Jacobian matrix with entries $\partial h_{{\rm so},i}({\bm k}^c)/\partial k_j$, reflecting the SO field configuration around ${\bm k}^c$.
Its determinant gives the charge value~\cite{Zhanglong2019a}
\begin{align}~\label{CnJso}
{\cal C}={\rm sgn}[J_{{\bm h}_{\rm so}}({\bm k}^c)],
\end{align}
where $J_{{\bm h}_{\rm so}}({\bm k})\equiv \det\left[(\partial h_{{\rm so},i}/\partial k_j)\right]$ is Jacobian determinant.

The Floquet bulk topology, characterized by the Floquet Hamiltonian $H_F\equiv\ui\log U(T)/T$, can be described in a similar way, while the major difference is that the periodic driving may lead to new BISs~\cite{Zhanglong2020}.
Here $U(t)={\cal T}\exp\big[-\ui\int_{0}^t H(\tau)d\tau\big]$ is the time evolution operator with ${\cal T}$ denoting the time ordering,
and the eigenvalues of $H_F$ defines the Floquet bands with two types of quasienergy gaps~\cite{Rudner2020}.
For driven systems described by Eq.~\eqref{FHam0}, the Floquet Hamiltonian takes a new Dirac-type form
$H_F({\bm k})=\sum_{i=0}^{d}h_{F,i}({\bm k})\gamma_{i}$, with the $\gamma$ matrices
being those in $H_s$. The BISs and topological charges are now determined respectively by $h_{F,0}({\bm k})$ and ${\bm h}^F_{\mathrm{so}}({\bm k})\equiv(h_{F,1},\dots,h_{F,d})$.
For the driving field applied to the $\gamma_0$ axis, topological charges are immune to the drive (${\bm h}^F_{\mathrm{so}}\sim {\bm h}_{\mathrm{so}}$ near ${\bm k}^c$), but the driving-induced new BISs can bring about new Floquet topology.
We denote the gap around the quasienergy $0$ ($\pi/T$) as $0$-gap ($\pi$-gap),
and the BIS living in this gap as $0$-BIS ($\pi$-BIS).
The topology of the Floquet bands below the $0$-gap, characterized by $W$, is contributed from all the $0$- and $\pi$-BISs~\cite{Zhanglong2020}:
\begin{align}\label{W_nu}
W=W_0-W_\pi,\quad W_q=\sum_{j\in q\textrm{-BIS}}\nu_{j}.
\end{align}
Here $\nu_j$ denotes the topological invariant associated with the $j$-th BIS, equivalent to the total charges enclosed.
We shall show that more than identifying the global bulk topology, the local topology on each BIS in fact has a unique connection to the gapless modes on the boundary, as studied below.

\section{BIS-boundary correspondence}~\label{sec3}

Here we demonstrate that a particular local topological structure (dubbed ``$\phi$-structure'', see Fig.~\ref{fig1} and also Appendix~\ref{app1_new} for more details) formed in each BIS
uniquely determines the number and features of boundary modes. We will first deal with the static system with $d\geq2$. Our proof is based on the BIS characterization~\cite{Zhanglin2018,Zhanglong2019a,YuJiZhang2021,Hu2020b,LiZhuGong2021} together with a dimension reduction approach. We split the momentum into $(k_\perp, {\bm k}_\parallel)$, with the component $k_\perp$ (${\bm k}_\parallel$) being perpendicular (parallel) to the surface or edge. The $d$D system is then reduced to a set of 1D Hamiltonians $H_{\rm 1D}(k_\perp)$ parameterized by ${\bm k}_\parallel$~\cite{Ryu2002,Mong2011,Delplace2011}. 
The key feature is that when the 1D momentum $k_\perp$-line crosses a local $(d-1)$D BIS with enclosed topological charges, the 1D system can host edge states.
For convenience, we assume that ${\bm h}_{\mathrm{so}}({\bm k})$ is linear when approaching a charge at ${\bm k}={\bm k}^c$, which restricts the charge value $|{\cal C}|=1$. 
The proof can be directly extended to larger-value charge that can be decomposed into multiple unit charges~\cite{Zhanglong2019a}.

\subsection{Static systems}

We consider the chiral symmetric quasi-1D system in the $k_\perp$ direction at ${\bm k}_\parallel={\bm k}_\parallel^c$,
and write
\begin{equation}~\label{1DHam}
H_{\rm 1D}(k_\perp)={\bm h}(k_\perp)\cdot{\bm\gamma}
=h_0(k_\perp)\gamma_0+h_l(k_\perp)\gamma_{l}.
\end{equation}
Here $\gamma_l$ is generally a combination as $\gamma_l=\sum_{i=1}^dc_{i}\gamma_{i}$~\cite{note} and
the chiral symmetry is $S=\ui\gamma_0\gamma_l$.

\begin{figure}
\includegraphics[width=0.49\textwidth]{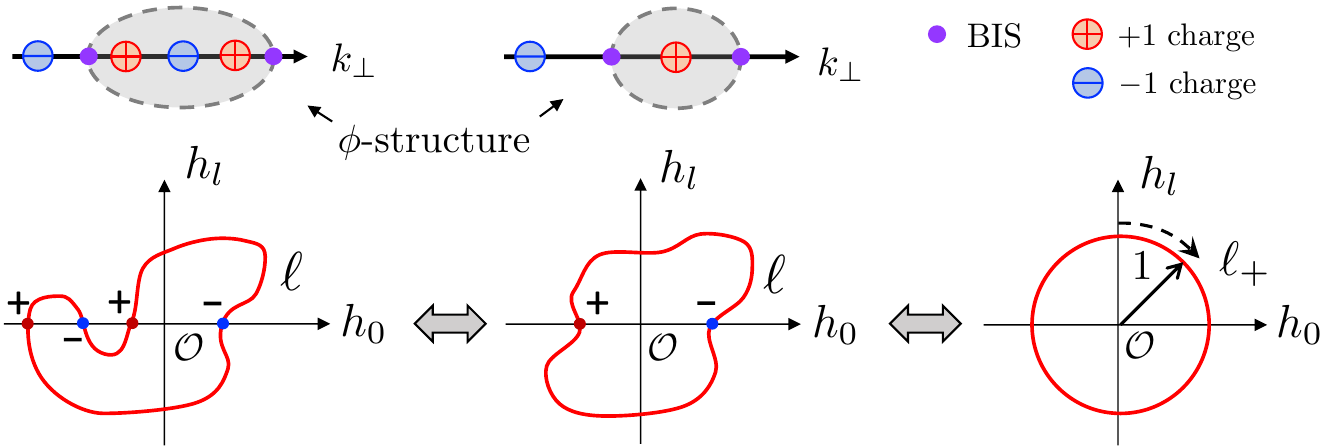}
\caption{Local topological structures and continuous deformation. Upper panel: Two examples of local $\phi$-structures that support boundary modes. Lower panel: The corresponding loops traced out by the vector field ${\bm h}(k_\perp)$. Both can be continuously deformed into a clockwise unit circle $\ell_{+}$ (rightmost).
Here topological charges correspond to the intersections of the loop $\ell$ with the $h_0$ axis, and the {\it projective} values $C_{\perp}=\pm1$
indicates the orientation ``$\pm$'' of the loop at intersections~\cite{Zhanglong2019a}. The shaded region denotes the region where $h_0<0$.
}\label{fig1}
\end{figure}

We start with the simple case that at ${\bm k}_\parallel={\bm k}_\parallel^c$, there is only one $\phi$-structure in the $k_\perp$ direction.
Here, a ``$\phi$-structure'' is defined as a particular topological structure
that the 1D $k_\perp$ line crosses a local $(d-1)$D BIS at two 0D momentum points and goes through at least one topological charge enclosed by the BIS (see Fig.~\ref{fig1} and Appendix~\ref{app1_new}).
The proof consists of two steps. First,
we verify that the topological $\phi$-structure gives rise to zero-energy edge states at ${\bm k}_\parallel^c$.
With nonzero net charge enclosed, the vector ${\bm h}(k_\perp)$ traces out a closed loop $\ell$ that surrounds the origin ${\cal O}$, which can be continuously deformed into a clockwise ($\ell_{+}$) or anticlockwise ($\ell_{-}$) unit circle, denoted as $\ell\sim\ell_{\pm}$ (see Fig.~\ref{fig1}). The unit circles traced out by ${\bm h}(k_\perp)=(h_0,h_l)=(-\cos k_\perp,\pm\sin k_\perp)$ has nonzero winding equal to the total enclosed charge with {\it projective} values ${\cal C}_{\perp}\equiv{\rm sgn}\left(\partial h_l/\partial k_\perp\vert_{k_\perp=k_\perp^c}\right)=\pm1$.
This topological number corresponds to a zero-energy state localized in each open boundary~\cite{Ryu2002}, as detailed in Appendix~\ref{app1}.
Second, we consider the neighboring region of ${\bm k}_\parallel={\bm k}_\parallel^c$ and show the existence of gapless boundary modes.
Denote by $H_{\rm 1D}^\delta(k_\perp)$
a family of 1D Hamiltonians at ${\bm k}_\parallel={\bm k}_\parallel^c+\delta{\bm k}_\parallel$, with $\delta{\bm k}_\parallel$ being a small shift, and write $H_{\rm 1D}^\delta(k_\perp)\simeq H_{\rm 1D}(k_\perp)+\sum_{i\neq0,l} h_{i}({\bm k}_\parallel)\gamma_i$.
Since the zero-energy edge states are in the eigenspaces of the chiral symmetry $S$ (Appendix~\ref{app1}) and $[\gamma_{i}, S]=0$ ($i\neq0,l$),
the projection operators ${P}^{(\pm)}=(1\pm S)/2$ can be introduced~\cite{Chiu2013},
which yields the effective Hamiltonian for the gapless boundary mode (see the details in Appendix~\ref{app2})
\begin{equation}~\label{Hboundary}
H_{\rm boundary}^{(\pm)}=\sum_{i\neq 0,l} h_{i}({\bm k}_\parallel) P^{(\pm)}\gamma_iP^{(\pm)}\simeq \left[v({\bm k}_{\parallel}^c)\delta{\bm k}_\parallel\right]\cdot {\bm\gamma}_P^{(\pm)},
\end{equation}
where $v({\bm k}_{\parallel}^c)$ is a matrix with elements $v_{ij}=\partial h_i/\partial k_{\parallel,j}\vert_{{\bm k}_\parallel={\bm k}_\parallel^c}$ and ${\bm\gamma}_P^{(\pm)}\equiv P^{(\pm)}{\bm\gamma}P^{(\pm)}$ with the superscripts ``$\pm$'' indicating the $S=\pm1$ eigenspaces, respectively. The chirality or spin-momentum locking feature of the boundary modes, as characterized in Eq.~\eqref{Hboundary}, is
reflected in $v({\bm k}_{\parallel}^c)$ and determined by the enclosed charge at ${\bm k}_\parallel^c$  [cf. Eq.~\eqref{CnJso}].
In particular, for 2D systems we have $H_{\rm boundary}^{(\pm)}=\pm h_{S}(k_\parallel)P^{(\pm)}$,
where $h_{S}$ is the component in the chiral-symmetry axis.
The ``$\pm$'' sign depends on the left- or right-hand edge and directly reflects $C_\perp$ (Appendix~\ref{app2}).
The chirality of edge modes manifests itself in the uni-directional propagation, given by
\begin{align}
{\rm Chirality}\equiv{\rm sgn}\left(\frac{\partial E_{\rm left}}{\partial k_\parallel}\bigg\vert_{k_\parallel=k_\parallel^c}\right)={\cal C},
\end{align}
with $E_{\rm left}$ denoting the energy of the left edge state, ${\cal C}\equiv{\rm sgn}\left[\frac{\partial(h_l,h_S)}{\partial(k_\perp,k_{\parallel})}\big\vert_{{\bm k}={\bm k}^c}\right]$ being the net enclosed charge at ${\bm k}_\parallel={\bm k}_\parallel^c$.

The above results can be naturally generalized to the case that there are $n$ numbers of topological $\phi$-structures crossed by the 1D $k_\perp$ line, denoted as $\ell\sim\ell_\pm^n$, where $\ell_\pm^n$ is a unit circle that winds around the origin $n$ times (each $\phi$-structure encloses a net unit charge). Then the $\bm h(k_\perp)$ vector has a winding number $\pm n$ and leads to $n$ zero modes in each boundary. 
Note that in this work we shall focus on the case that the gapless boundary modes at the same ${\bm k}_\parallel^c$ have the same chirality, while an example of gapless boundary modes with opposite chirality at the same ${\bm k}_\parallel^c$ will be discussed in Sec.~\ref{zigzag}.

\subsection{Floquet systems}

\begin{figure}[t]
\includegraphics[width=0.485\textwidth]{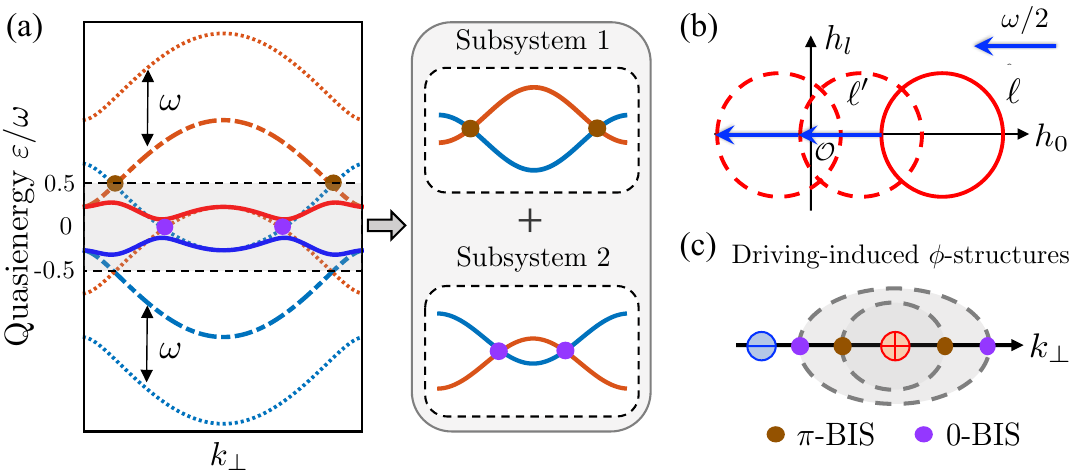}
\caption{Floquet band structure and loop representation.
(a) Left panel: Floquet bands (solid curves) are generated via gap openings at driving-induced BISs (dots).
Here BISs are formed by the static $h_0$ bands (dash-dotted curves) and their shifted copies (dotted curves).
The shaded region denotes the Floquet Brillouin zone $[-\pi/T,\pi/T]$.
Right panel: A Floquet system can reduce to a family of subsystems, each consisting of the bands that define a BIS.
(b) The applied driving shifts the loop for static bands (solid circle) along the $h_0$ axis in steps of $\omega/2$.
A shifted loop (dashed circle) that surrounds the origin ${\cal O}$ corresponds to a subsystem that has driving-induced BISs.
(c) Each subsystem in (a,b) contributes to at least one nontrivial $\phi$-structure.
}\label{fig2}
\end{figure}

We now turn to Floquet systems described by Eq.~\eqref{FHam0} with $d\geq2$. We still consider that
${\bm h}(k_\perp)$ of the static part $H_s$ describes
a quasi-1D model taking the form of Eq.~\eqref{1DHam}. Thus,
${\bm h}(k_\perp)$ traces out a closed loop $\ell$ in a 2D plane that contains the origin ${\cal O}$.
We shall show that (i) the periodic driving applied to the $\gamma_0$ axis shifts the loop $\ell$ along the $h_0$ direction in steps of $\omega/2$ ($\omega=2\pi/T$ is the driving frequency) [see Fig.~\ref{fig2}(b)],
and (ii) when the shifted loop $\ell'$ satisfies $\ell'\sim\ell_\pm^n$,
the driving brings new BISs and each driving-induced $\phi$-structure corresponds to a gapless boundary mode in the associated quasienergy gap.


The BISs of Floquet systems can be described from the perspective of the quasienergy operator
$Q(t)\equiv{H}(t)-\ui\partial_t$, whose eigenvalues $\varepsilon$ form the quasienergy bands~\cite{Rudner2013,Eckardt2015,Zhanglong2020}.
In the bases $e^{\ui m\omega t}$ labeled by the integer $m$,
a periodic drive can be written as $V(t)=\sum_{m\neq0}V^{(m)} e^{\ui m\omega t}$,
and the $Q$ operator for Eq.~\eqref{FHam0} takes an infinite block form
\begin{align}
[Q]_{mm'}=\delta_{mm'} (H_s+m\omega{\id}) +V^{(m-m')}\gamma_0,
\end{align}
where $[Q]_{mm'}$ denotes the block in the basis $e^{\ui (m-m')\omega t}$, which has the same dimension as $H_s$.
The diagonal blocks $H_s +m\omega$ are the copies of the static Hamiltonian with energy shifts $m\omega$,
and the off-diagonal blocks $V^{(m-m')}\gamma_0$ couple the $m$th and $m'$th copies.
The quasienergy spectrum has the following intuitive picture:
Periodic driving transfers energy into the system in portions,
which causes the shifting of the $h_0$-band copies in steps of $\omega$ and leads to driving-induced BISs. The
finite driving strength and SO coupling terms
open gaps at these BISs, forming the Floquet bands characterized by $H_F$ [see Fig.~\ref{fig2}(a)].
With this picture the Floquet system can be treated as a family of static subsystems which are periodic in quasienergy, and
each subsystem consists of $n_d$ bands with BISs [see Fig.~\ref{fig2}(a)].

We then consider the quasi-1D $k_\perp$-space at ${\bm k}_\parallel={\bm k}_\parallel^c$
and examine the effect of periodic driving.
In loop representation,
the energy transfer can be regarded as dragging the loop $\ell$ for the static system along the $h_0$ direction step by step [see Fig.~\ref{fig2}(b)].
At each step, the shifted loop $\ell'$ represents a new $n_d$-band subsystem.
When $\ell'$ crosses the $h_l$ axis, new driving-induced BISs emerge and give rise to a nonzero contribution to the Floquet topology.
In particular, when $\ell'\sim\ell^n_\pm$, the subsystem exhibits $\phi$-structures for $0$ ($\pi$)-gap [Fig.~\ref{fig2}(c)].
Similar to the analysis on static systems, each $\phi$-structure supports a gapless boundary mode.

To illustrate the above argument, we employ the Floquet perturbation theory~\cite{Mukherjee2020,Haldar2021}
to derive the effective Hamiltonian for subsystems.
Without loss of generality, we assume $V(t)=2\sum_{j=1}^pV^{(j)} \cos(j\omega t)$, where $V^{(j)}$ are real and $p$ is a positive integer.
Consider the driving-induced BISs defined by $h_0({\bm k})=m\omega/2$ with $m>0$. The effective Hamiltonian for the corresponding subsystem is given by (see the details in Appendix~\ref{app3})
\begin{equation}~\label{Heff_sub}
H_{\rm eff}^{(m)}=\left(h_0-\frac{m\omega}{2}\right)\gamma_0+(-1)^m{\cal J}_m\sum_{i>1}h_i\gamma_i,
\end{equation}
where
\begin{equation}
{\cal J}_m\equiv\sum_{\sum_{j=1}^pn_j=m}\prod_{j=1}^pJ_{n_j}\left(\frac{4V^{(j)}}{j\omega}\right),
\end{equation}
with $J_n(z)$ being the Bessel function and $n_1,n_2,\dots,n_p\in\{0,1,2,\dots,m\}$.
Supposing ${\cal J}_m>0$, one has that when $m$ is even (corresponding to 0-BISs),
the effective Hamiltonian is equivalent to the static one but has a shifted Zeeman energy, while when $m$ is odd ($\pi$-BISs),
the SO coupling of the subsystem is also reversed. For all $m$, their loops wind around the origin the same times.
The gapless boundary mode supported by the $\phi$-structure in this subsystem can then be described by
\begin{equation}~\label{Hboundary_F}
H_{\rm boundary}^{(\pm)}=(-1)^m{\cal J}_m\sum_{i\neq 0,l} h_{i}({\bm k}_\parallel) {\bm\gamma}_P^{(\pm)}.
\end{equation}

Before proceeding, we make remarks on several key points of the BIS-boundary correspondence.
First, the $\phi$-structure, formed by a quasi-1D system crossing at two $0$D momentum points of a $(d-1)$D BIS, ensures the existence of zero-energy edge states, and also determines the chirality or spin-momentum locked textures of the corresponding $(d-1)$D boundary modes from the net enclosed charge.
Second, the range of ${\bm k}_\parallel$ where the gapless boundary modes reside is given by the projected extent of the $(d-1)$D BIS.
The two facts, applicable to both static and Floquet systems, manifest
the one-to-one correspondence between the $(d-1)$D BISs and the $(d-1)$D boundary modes, and therefore can map the detection of boundary modes to measuring the topological information of $(d-1)$D BISs in the first Brillouin zone.
Besides, we would like to emphasize that in our demonstration there are two underlying assumptions: (i) The static systems should be gapped topological phases classified by integer invariants,
whose Hamiltonians in Eq.~\eqref{FHam0} are written in the elementary representation matrices satisfying Clifford algebra relations.
(ii) The quasi-1D system at ${\bm k}_\parallel={\bm k}_\parallel^c$ should have a chiral symmetry.
Now that BISs serve as the fundamental and concise element to characterize Floquet topology, one can reach a systematic scheme to realize and manipulate 
new topological states by engineering the low-dimensional BIS configuration.

\section{Unconventional Floquet topological phases}~\label{sec4}

\subsection{Engineering band inversion surfaces}

With the BIS-boundary correspondence, we now propose a systematic and physically intuitive scheme to
realize novel Floquet topological phases through quantum engineering of the BIS configuration.
As described above, applying the periodic drive can induce new BISs for the $d$D bands.
Whenever an induced BIS forms a new $\phi$-structure, the system enters a different Floquet topological regime
which is precisely characterized by all such formed $\phi$-structures.
Hence, one can realize under full control various Floquet topological phases by regulating, e.g., the driving frequency to generate
distinct BIS configurations, and characterize the phases via all $\phi$-structures formed in these BISs.
Since this BIS engineering approach does not involve details of complex temporal evolution,
it provide a minimal but powerful tool to realize, manipulate, and detect anomalous Floquet topological phases. 

A convenient way of realizing various Floquet topological phases is to apply a periodic drive on top of $d$D gapped topological bands, and
alter the Floquet band structure by gradually lowering the driving frequency.
Consider SO coupled topological systems described by Eq.~\eqref{FHam0}, for which static BISs and topological charges can
form at least one $\phi$-structure in certain momentum directions. The applied $\gamma_0$-axis driving starts to generate new BISs
when the frequency is lowered to a critical value.
In such case, a new BIS always gives rise to a new $\phi$-structure and
the induced BISs appear in the 0-gap and $\pi$-gap alternately
(see concrete examples below).
With these features, one can monitor the emergent sequence of BISs and engineer the Floquet topological bands with purpose.

We emphasize the notable difference from conventional characterization theories that the BIS-boundary correspondence
employed here is a local correspondence in momentum space.
This feature has an importance consequence in realizing unconventional states.
In particular, when more than one BIS emerges in the same quasienergy $0$ (or $\pi$)-gap,
while residing in different local momentum regions and contributing oppositely to the bulk topology,
the system may enter an unconventional Floquet topological phase with multiple boundary modes,
for which the conventional global topological invariants are not sufficient for a full characterization. 
Below we illustrate it with two concrete examples which are relevant for experiments.

\subsection{2D anomalous valley-Hall phase}

\begin{figure}
\includegraphics[width=0.48\textwidth]{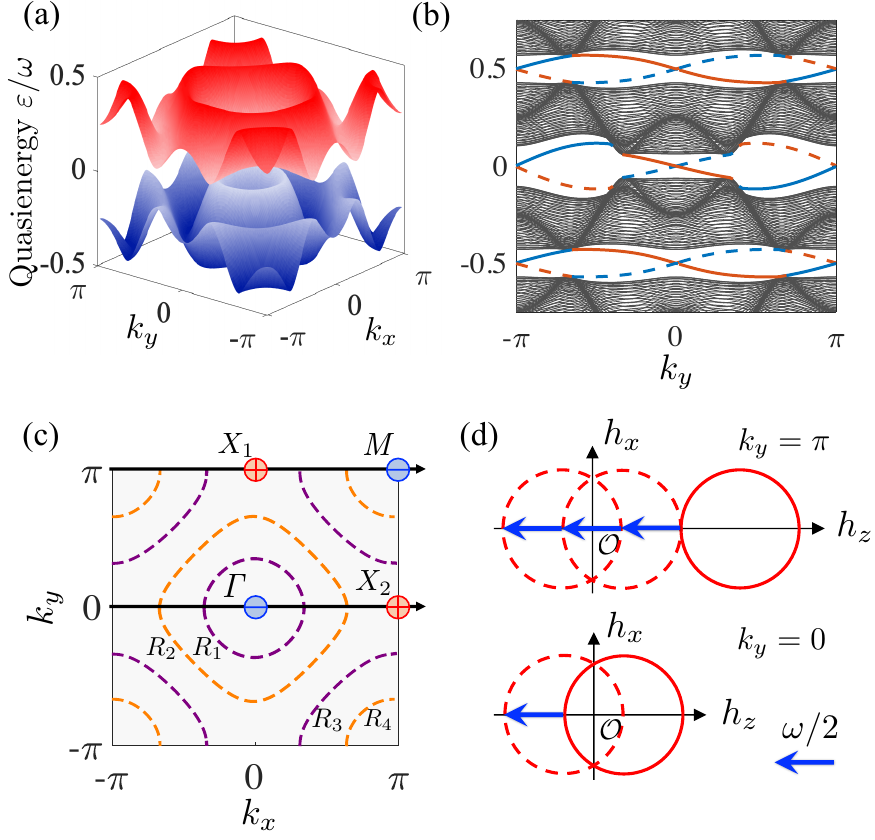}
\caption{Anomalous valley-Hall phase and BIS-boundary correspondence in the 2D driven model.
(a-b) Quasienergy spectrum
with periodic boundary conditions (a) or for a cylindrical geometry (b).
In (b), edge modes at the left (right) boundary are plotted as solid (dashed) curves,
and colored in dark orange (blue) for the negative (positive) chirality.
(c) Driving-induced topological structures in the Brillouin zone.
Four charges are located at $\it\Gamma$, $X_{1,2}$ and $M$ points, with the values given by ${\cal C}={\rm sgn}\big[\frac{\partial(h_x,-h_y)}{\partial(k_x,k_y)}\big\vert_{{\bm k}={\bm k}^c}\big]$.
Each $0$-BIS (purple dashed ring) or $\pi$-BIS (orange dashed ring) forms a $\phi$-structure in the direction $k_y=0$ or $\pi$.
(d) Loops for the local topological structures.
Here $t_{\rm so}=0.5t_0$, $m_z=3t_0$, $\omega=4t_0$ and $V_0=3t_0$.
}\label{fig3}
\end{figure}

We consider a 2D driven model taking the form of Eq.~\eqref{FHam0}:
\begin{align}~\label{FHam2D}
\begin{split}
&H({\bm k},t)=H_{\rm s}({\bm k})+V(t)\sigma_z,\\
&H_{\rm s}({\bm k})={\bm h}({\bm k})\cdot{\bm\sigma},\quad V(t)=2V_0\cos\omega t,
\end{split}
\end{align}
where $\sigma_{x,y,z}$ are the Pauli matrices.
Here ${\bm h}({\bm k})=(2t_{\rm so}\sin k_x,2t_{\rm so}\sin k_y,m_z-2t_0\cos k_x-2t_0\cos k_y)$ describes a quantum anomalous Hall model~\cite{LiuXJ2014},
which has been realized in cold atoms, with $t_0$ ($t_{\rm so}$) being the spin-conserved (-flipped) hopping in a square optical Raman lattice~\cite{Wu2016,Sun2017,Liang2021}.
The periodic drive $V(t)$ can be achieved by modulating the Zeeman energy (see Sec.~\ref{realization2D} for details).
According to the above generic theory, for this 2D model, an unconventional topological phase emerges whenever
there are two 0 ($\pi$)-BISs circling different topological charges, which can be easily engineered (Sec.~\ref{realization2D}).
Here we show a highly nontrivial one called the anomalous valley-Hall phase.
Fig.~\ref{fig3}(a) and (b) display the quasienergy spectra for different boundary conditions.
According to the band structure in (a), there are two 0-BISs ($R_{1,3}$) and two $\pi$-BISs ($R_{2,4}$) in the first Brillouin zone [see Fig.~\ref{fig3}(c)]: One ($R_1$) is formed by the static bands and three ($R_{2,3,4}$) are induced by the driving.
Each BIS contributes to a local $\phi$-structure at $k_y=0$ or $\pi$, and corresponds to a chiral edge mode in the associated gap.

\subsubsection{Classification and characterization}

The above topological phase (Fig.~\ref{fig3}) realized by BIS manipulation is an exotic one that cannot be classified by conventional topological invariants defined for the global bulk.
The both winding numbers defined to characterize the topology of the 0-gap and $\pi$-gap vanish $W_{0,\pi}=0$, hence the Chern number ${\rm Ch}=W_0-W_{\pi}=0$ [Eq.~\eqref{W_nu}].
However, this phase features stable counterpropagating edge states in both quasienergy gaps [Fig.~\ref{fig3}(b)], even though the present 2D system has no time-reversal symmetry.
Just like the static valley-Hall phase,
the engineered bulk bands are topologically trivial, and
the edge modes are protected by local topological structures in two distinct Brillouin-zone regions (i.e. valleys) near the $\it\Gamma$ and $M$ points, and robust against long-range disorders~\cite{Ren2016_review}.
This renders an anomalous Floquet valley-Hall phase, which is beyond solid-state realizations~\cite{Xiao2007,Qiao2011,Xiao2012,Gorbachev2014,Mak2014} and the conventional characterization, but fully captured by the current BIS-boundary correspondence.

The BIS-boundary correspondence can be interpreted by loops as shown in Fig.~\ref{fig3}(d).
For the quasi-1D system at $k_y=0$, the loop traced out by the static field ${\bm h}(k_x)$ (solid circle) surrounds the origin ${\cal O}$,
corresponding to the $\phi$-structure that goes across the 0-BIS $R_1$ and through the charge at ${\it\Gamma}$.
Such a $\phi$-structure supports a gapless edge mode centered at $k_y=0$ in the $0$-gap. The applied driving shifts the static loop
along the $h_z$ direction in steps of $\omega/2$.
We note that the one-step-shifted loop (dashed circle) also surrounds ${\cal O}$, which indicates the emergence of the $\pi$-BIS $R_2$.
The driving-induced $\phi$-structure crossing $R_2$ generates an edge mode in the $\pi$-gap.
Since the enclosed charges are the same one,
the two edge modes around $k_y=0$ have the same negative chirality [Fig.~\ref{fig3}(b)].
In contrast, for the quasi-1D system at $k_y=\pi$, the two driving-induced $\phi$-structures correspond to two edge modes centered at $k_y=\pi$ with positive chirality, with one residing in the 0-gap and the other in the $\pi$-gap.
Note that although each BIS has non-trivial topology, the contributions from the 0- and $\pi$-BISs in the same valley ($R_{1,2}$ or $R_{3,4}$) cancel each other, leading to trivial valley topology. Hence, a valley-Hall effect can only be defined inside every single gap, which makes the anomalous Floquet valley-Hall phase a non-trivial generalization of its static counterpart.

Besides, the BIS-boundary correspondence also benefits the experimental detection of Floquet topological phases.
Once the BIS configuration is detected in experiment, one can directly determine which quasienergy gap
and what momentum region the edge modes live in, the latter corresponding to the projected extent
of BIS onto the ${\bm k}_\parallel$ [$k_y$ in Fig.~\ref{fig3}(b)] direction.
In this way, all the engineered Floquet topological phases can be precisely identified in experiment.


Here we would like to emphasize that
our predicted anomalous valley-Hall phase is said to be beyond the conventional characterization in two senses.
(i) Unlike the static valley-Hall phase, where every whole valley acts as a degree of freedom~\cite{Ren2016_review},
the anomalous Floquet valley-Hall phase has no such valley index, since not only the bulk but also every valley
is topologically trivial. 
Therefore, neither a global topological invariant (the Chern number) nor a valley topological number (the valley Chern number)
can characterize this unconventional phase.
(ii) Unlike other anomalous Floquet phases~\cite{Rudner2013,Nathan2015}, the counter-propagating edge states predicted here are not protected by the gap topology;
the two winding numbers $W_{0,\pi}$ both vanish due to the opposite chirality of the two edge modes inside the same gap.
In the 0-gap ($\pi$-gap), the edge modes are formed within local regions surrounded by 0-BISs  ($\pi$-BISs) and protected by corresponding local topological structures.
These edge modes are then robust against long-range disorders that do not induce inter-valley scattering (see Fig.~\ref{fig5}), similar to the static valley Hall phase.

\subsubsection{Edge-geometry dependence and disorder robustness}~\label{zigzag}

\begin{figure}
\includegraphics[width=0.48\textwidth]{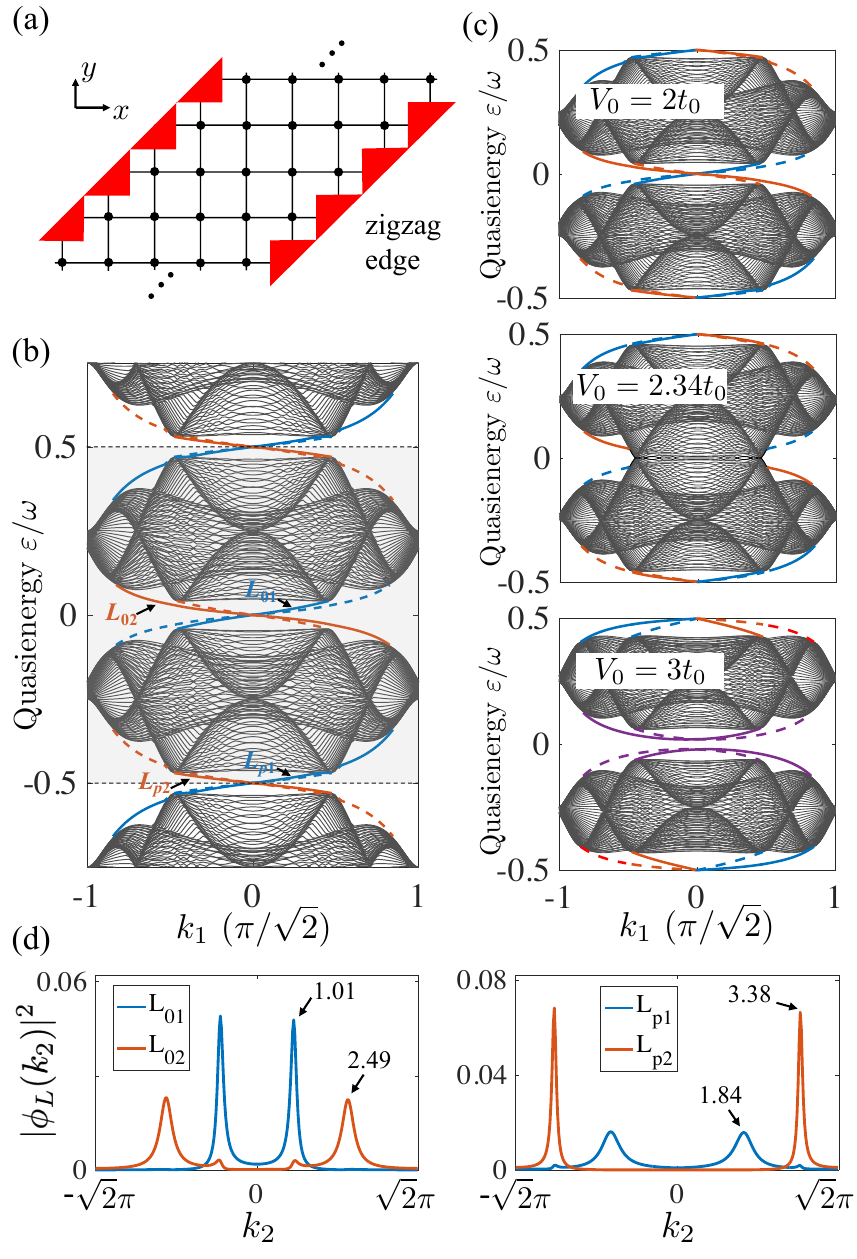}
\caption{
The 2D anomalous valley-Hall phase in a zigzag-edge geometry.
(a) A sketch of the zigzag edge geometry with two boundaries in the $x=y$ direction.
(b) An example of the quasienergy spectrum for the edge geometry in (a).
The counter-propagating edge modes survive in both quasienergy gaps.
Here we set $V_0=2t_0$. The shaded region denotes the Floquet Brillouin zone. Edge modes at the left (right) boundary
are plotted as solid (dashed) curves, and colored in dark orange (blue) for the negative (positive)
chirality.
(c) The phase transition of 0-gap topology from gapless (the upper) to gapped (the lower) edge modes induced by driving strength,
with a gap-closing occurring at a critical value $V_0=2.34t_0$ (the middle). Other parameters in (b,c) are taken the same as in Fig.~\ref{fig3}.
(d) The momentum distribution of the left edge modes labeled in (b).
A detailed analysis on the peaks is put in Appendix~\ref{app_new}.
}\label{fig4}
\end{figure}

Our predicted anomalous Floquet valley-Hall phase exhibits edge-geometry dependence,
which has similarity as the static valley-Hall effect~\cite{Ren2016_review} but also exhibits novel new features, as studied below.
In particular, we identify a driving-induced topological phase transition occurring only for zigzag edges.
The zigzag edge geometry is illustrated in Fig.~\ref{fig4}(a), where
the 2D driven model lives on a square lattice in a diagonal strip~\cite{Mao2010}.
Here we define $k_{1,2}=(k_x\pm k_y)/\sqrt{2}$.
The translational invariance is preserved in the $x=y$ direction,
yielding the conserved momentum $k_1$.
We calculate the quasienergy spectra for such a geometry and display the results in Fig.~\ref{fig4}(b,c) as a function of $k_1$, where
we fix the driving frequency $\omega$ and choose the strength $V_0$ as a tunable parameter.
We find that when the driving is relatively weak, the counter-propagating edge modes are obtained in each quasienergy gap [Fig.~\ref{fig4}(b) and the top one in (c)].
Interestingly, while the forward and backward edge modes are centered at the same $k_1$ (along the edge direction), they are formed by far-separated quasi-momentum components  ($k_2$) along the direction perpendicular to the edge [Fig.~\ref{fig4}(d)].
One can derive that the wave function of the left edge state takes a form (see Appendix~\ref{app_new})
\begin{align}~\label{phiL_k2}
\phi_{L}(r_\perp)\sim\left(c_1 e^{\ui k_{2}^{+} r_\perp} +c_2 e^{\ui k_{2}^{-} r_\perp}\right)e^{-t_{\rm so}r_\perp/t_0},
\end{align}
where $r_\perp$ is the spatial direction perpendicular to the edge, and the characteristic momenta $k_2^{\pm}$ are determined by the corresponding valley.
This implies that the counter-propagating edge modes actually belong to different valleys, and should be also robust to the long-range disorders.
With the increase of $V_0$, the edge modes can be gapped and the valley-Hall effect no longer exists in both gaps [the lowest in Fig.~\ref{fig4}(c)].
There occurs a gap-closing approximately at $V_0\approx 0.585\omega$ for the parameters chosen here (the middle).
Note that the gapped edge states are absent for the straight edge geometry in Fig.~\ref{fig3}(b).

The observed gap-closing can be well explained using perturbation theory analysis.
From the result of Eq.~\eqref{Heff_sub}, we have the effective Hamiltonian on BISs of the 2D driven model
\begin{equation}~\label{Heff_sub_2d}
H_{\rm eff}^{(m)}=\left(h_z-\frac{m\omega}{2}\right)\sigma_z+(-1)^mJ_m\left(\frac{4V_0}{\omega}\right)(h_x\sigma_x+h_y\sigma_y),
\end{equation}
where the functions $J_m\left(\frac{4V_0}{\omega}\right)$ are all positive when $V_0$ is much smaller than $\omega$ and can turn to be negative for large $V_0$.
For the case in Fig.~\ref{fig4}, we only need to consider the function of order $m=0$, for which
the turning from positive to negative is at $V_0\approx 0.6\omega$.
This characterizes the vanishing of the decorated SO coupling on the static BIS and corresponds to the observed gap-closing.
The estimated value of the transition point is very close to the numerical result. The slight difference is due to a finite value of $t_{\rm so}$ in numerical calculations.

The topological phase transition from gapless to gapped edge modes can be interpreted by local topological structures.
Consider the quasi-1D system at $k_1=0$, whose Hamiltonian in $k_1$-$k_2$ coordinates is given by $H_{\rm 1D}(k_2,t)=H_s(k_1=0, k_2)+V(t)\sigma_z$.
We introduce $\sigma_{X,Y}=(\sigma_x\pm\sigma_y)/\sqrt{2}$ and have
$H_{\rm 1D}(k_2,t)=[m_z-4t_0\cos(k_2/\sqrt{2})+2V_0\cos\omega t]\sigma_z
+2\sqrt{2}t_{\rm so}\sin(k_2/\sqrt{2})\sigma_Y$ (Appendix~\ref{app_new}),
which respects a chiral symmetry $S=\ui\sigma_z\sigma_Y$, i.e., $SH_{\rm 1D}(k_2,t)S^{-1}=-H_{\rm 1D}(k_2,-t)$.
It can be expected that the Floquet Hamiltonian of this quasi-1D system maintains the local chiral symmetry $S$:  $SH_{F}(k_2)S^{-1}=-H_{F}(k_2)$,
and thus takes the form $H_{F}(k_2)=h_{F,z}(k_2)\sigma_z+h_{F,Y}(k_2)\sigma_Y$.
At the momentum points where $h_{F,z}(k_2)=0$ (the crossing points of the $k_2$ line with BISs), we can write $h_{F,Y}\approx (-1)^mJ_m\left(\frac{4V_0}{\omega}\right)h_Y$ according to Eq.~\eqref{Heff_sub_2d}.
We focus on the topology inside 0-gap (even $m$).
We see that when $V_0<0.6\omega$, $h_{F,Y}\sim h_Y$ on both static ($m=0$) and driving-induced ($m=2$) BISs
and two local $\phi$-structures with the same reduced charge ${\cal C}_{\perp}$ are formed to support two zero-energy mid-gap states.
It implies that the edge modes are gapless in this phase.
In contrast, when $V_0$ is tuned to be a little larger than $0.6\omega$, $h_{F,Y}\sim -h_Y$ on the static BIS and $h_{F,Y}\sim h_Y$ on the one with $m=2$ .
There emerge two local $\phi$-structures with opposite reduced charges; the topological invariants on the two BISs are $\pm1$,
leading to trivial 0-gap topology. 

\begin{figure}[t]
\includegraphics[width=0.49\textwidth]{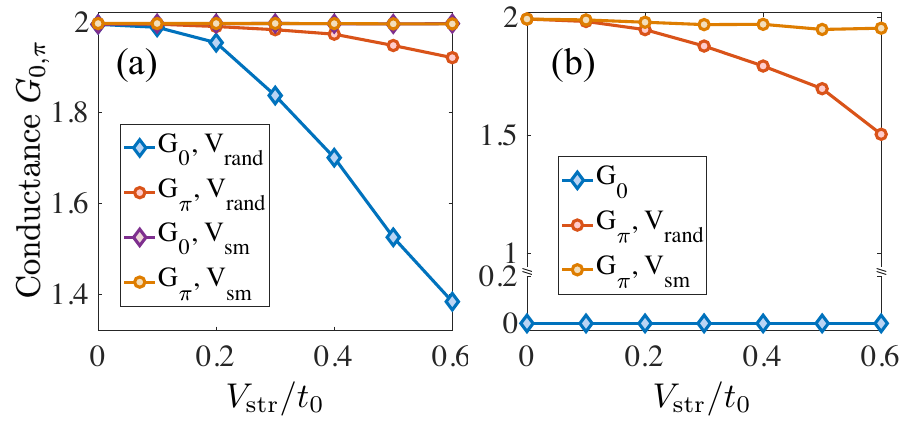}
\caption{
The computed conductance within each quasienergy gap in the presence of disorders for a straight (a) or zigzag (b) edge geometry.
Here (a) corresponds to the case in Fig.~\ref{fig3}(b) while (b) is for the lowest panel in Fig.~\ref{fig4}(c) with trivial 0-gap topology.
In our calculations, we set the lattice size as $40\times40$ and take the impurity number $N_{\rm imp}=50$
for the smooth impurity potential $V_{\rm sm}$.
Each data is averaged over at least 50 disorder configurations.
}\label{fig5}
\end{figure}

Finally, we employ scattering matrix invariants~\cite{Fulga2016,Umer2020} to characterize topological phases and examine the disorder effects.
The quasienergy-dependent Floquet scattering matrix is expressed as
\begin{align}
S(\varepsilon)={P}\left[e^{-i\varepsilon}{\id}-U(1-P^{\rm T}{P})\right]^{-1}U{P}^{\rm T}=\left( \begin{array}{cc}
r & t \\
t' & r'\\
\end{array} \right),
\end{align}
where ${P}$ is the projector matrix (see Ref.~\cite{Fulga2016} for the definition), the superscript ${\rm T}$ denotes the matrix transpose, and $U$ is the
time-evolution operator governed by the Hamiltonian under open boundary conditions.
From this expression one can compute the conductance $G_{0,\pi}={\rm Tr}(tt^\dagger)$, which characterizes
the number of edge modes inside the corresponding quasienergy gap.
We consider two kinds of symmetry-preserving disorders applied to the Zeeman term: $m_z\to m_z+V_{\rm disorder}$.
The first one is the random on-site disorder
\begin{align}
V_{\rm rand}({\bm r})=\sum_{j\in{\rm site}} V_j\delta({\bm r}-{\bm r}_j),
\end{align}
and the second represents a long-range potential induced by smooth impurities
\begin{align}
V_{\rm sm}({\bm r})=\sum_l^{N_{\rm imp}}\frac{V_l}{\sqrt{({\bm r}-{\bm r}_l)^2+d^2}},
\end{align}
where $N_{\rm imp}$ denotes the number of the randomly distributed impurities
and $d$ controls the potential range.
Here we set $d=5$ and take $V_{\rm rand},V_{\rm sm}\in[-V_{\rm str},V_{\rm str}]$, with $V_{\rm str}$ denoting the disorder strength.
The numerical results are shown in Fig.~\ref{fig5} for the disorder effect on the 2D anomalous valley-Hall phase.
For the straight edge geometry, the conductance $G_{0,\pi}$ are both equal to 2 without disorders [Fig.~\ref{fig5}(a)].
When the on-site disorder $V_{\rm rand}({\bm r})$ is applied, $G_{0,\pi}$ gradually
deviate from their quantized values as the strength $V_{\rm str}$ increases,
indicating that the edge states with opposite chirality are hybridized with each other by disorders.
In comparison, the computed values of $G_{0,\pi}$ are almost unaffected by $V_{\rm sm}({\bm r})$,
which verifies that the counterpropagating edge states are immune against long-range disorders.
For the zigzag edge geometry [Fig.~\ref{fig5}(b)], we have vanishing $G_{0}$ as a signature of trivial 0-gap topology [cf. the lowest panel of Fig.~\ref{fig3}(c)].
It should be also noted that
the robustness of $\pi$-gap edge modes against smooth impurities is preserved for zigzag edges, which demonstrate the analysis on the large
$k_2$-component separation in Eq.~\eqref{phiL_k2} and Fig.~\ref{fig4}(d).

\subsubsection{Experimental realization}~\label{realization2D}
Now we study how to experimentally realize the predicted anomalous Floquet valley-Hall phase based on the BIS engineering.
As a local, intuitive, and minimal strategy for realizing unconventional topological states, the BIS engineering is experimentally feasible 
and can be well achieved based on current experiments. 
Here we consider a realistic cold-atom setup and describe in detail how to simulate the anomalous Floquet valley-Hall phase~\cite{Exp2022}.

The experimental setup we consider is ultracold $^{87}$Rb atoms trapped in a shaking square optical Raman lattice, where
a pair of standing-wave beams are applied to couple two hyperfine states, e.g., $\mid\uparrow\rangle\equiv|F=1,m_F=-1\rangle$ and $\mid\downarrow\rangle\equiv|1,0\rangle$,
and generate the required 2D spin-orbit coupling via two Raman processes which form a double-$\Lambda$-type coupling configuration~\cite{LiuXJ2014,Wu2016,Sun2017,Liang2021}.
The two-photon detuning of the Raman coupling can be tuned in a time-periodic form $\delta(t)=\delta_0+2\Delta\sin(\omega t)$ to drive the system by modulating the laser frequency.
Such a system is described by
\begin{align}~\label{FHam_exp}
\begin{split}
&H(t)=\frac{{\bm k}^2}{2m}+V_{\rm latt}({\bm r})+{\cal M}_x({\bm r})\sigma_x+{\cal M}_y({\bm r})\sigma_y+\frac{\delta(t)}{2}\sigma_z,
\end{split}
\end{align}
where $V_{\rm latt}({\bm r})=V_{L}[\cos^2(k_0x)+\cos^2(k_0y)]$ denotes the square lattice potential ($k_0$ is the wave vector), and
${\cal M}_{x}({\bm r})=M_R\sin(k_0x)\cos(k_0y)$ and ${\cal M}_{y}({\bm r})=M_R\cos(k_0x)\sin(k_0y)$ are the Raman potentials.
This time-dependent Hamiltonian well simulates the 2D driven model in Eq.~\eqref{FHam2D} in the tight-binding approximation~\cite{LiuXJ2014,WangBZ2018}.
As previously proposed, one can synthesize unconventional Floquet topological bands by systematically engineering the BIS configuration.
For instance, one can start the periodic drive $\delta(t)$ at a very high frequency and gradually lower the driving frequency. In this way,
the BISs are generated one by one, and for this system the 0-BIS and $\pi$-BIS will appear alternately.
The frequency regulation is terminated when the target four-ring configuration, as a key indication of the anomalous valley-Hall phase shown in Fig.~\ref{fig3}, is produced.
The typical parameters for the case in Fig.~\ref{fig3} can be achieved in experiment by taking $V_L=4E_{\rm r}$, $M_R=0.63E_{\rm r}$, the constant detuning $\delta_0=0.5E_{\rm r}$,
and the driving frequency $\omega=0.33E_{\rm r}$. Here $E_{\rm r}\equiv k_0^2/(2m)$ is the recoil energy.

For the cold atom experiment, the engineered Floquet band topology can be detected by quantum quench dynamics in three steps as follows:
(i) $^{87}$Rb atoms are initially prepared in a single spin state by a very large two-photon detuning.
A proper temperature should be set such that the bosons can nearly occupy the whole lowest band.
(ii) The detuning is quenched to the value $\delta_0$ within a very short time (of $\mu$s) and then modulated at the desired frequency $\omega$.
The initial state then evolves under the time-periodic Hamiltonian given in Eq.~\eqref{FHam_exp}.
(iii) The time evolution of the spin texture is observed and recorded. All the BISs (both static and driving-induced ones)
can be identified as the momenta where spin-flip significantly occurs due to the resonant coupling on the BISs.
After the procedure, different topological regimes can be experimentally distinguished by distinct BIS configurations.
In particular, our predicted anomalous Floquet valley-Hall phase is characterized by a spin texture with a four-ring structure~\cite{Exp2022}. This experimental procedure is well applicable to various cold atom candidates including both bosons and fermions.

\subsection{3D anomalous chiral phase}

\begin{figure}[t]
\includegraphics[width=0.49\textwidth]{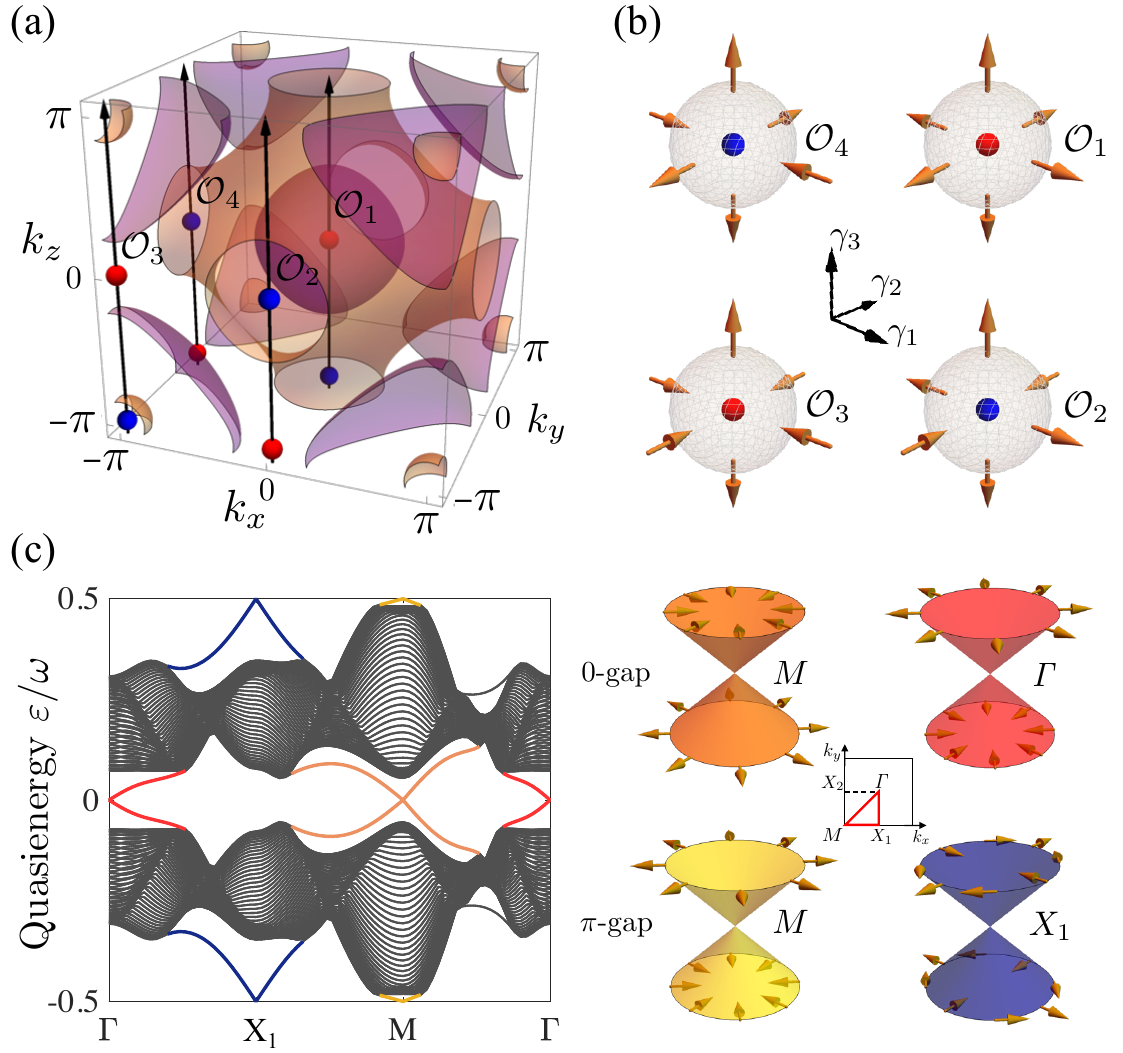}
\caption{Anomalous chiral phase and BIS-boundary correspondence in the 3D driven model.
(a) Two $0$-BISs (purple) and two $\pi$-BISs (orange) surrounding topological charges with ${\cal C}=+1$ (red) or $-1$ (blue) form
 local topological $\phi$-structures (black arrows) in the $k_z$ direction.
(b) SO field configurations around topological charges ${\cal O}_{1,2,3,4}$.
(c) Energy spectrum with open boundary condition in $z$-direction (left panel) and spin textures of the surface Dirac cones (right panel).
Here we set $t_{\rm so}=t_0$, $m_0=4t_0$, $\omega=6.5t_0$ and $V_0=3t_0$.
}\label{fig6}
\end{figure}

We further apply the BIS engineering to realize a 3D anomalous chiral topological phase characterized by local topological structures in the driven model with
\begin{align}~\label{FHam3D}
\begin{split}
&h_0=m_0-2t_0\sum_{i=1}^3\cos k_{r_i},\\
&h_{i>0}=2t_{\rm so}\sin k_{r_i}, \quad V(t)=2V_0\cos\omega t.
\end{split}
\end{align}
Here $(r_1,r_2,r_3)\equiv(x,y,z)$.
The $\gamma$ matrices can be constructed as the tensor product of the Pauli matrices~\cite{Zhanglin2018,Zhanglong2019a}.
The static Hamiltonian gives a 3D chiral topological insulator recently simulated in experiment using solid-state spin systems~\cite{Ji2020,Xin2020}, which has eight topological charges shown in Fig.~\ref{fig6}(a) and (b).
Similar to the 2D case, unconventional topological phases are realized by inducing at least two BISs of the same (0 or $\pi$) type that enclose opposite-sign net charges and contribute oppositely to the bulk topology.
Taking $m_0=4t_0$ and $\omega=6.5t_0$,
there emerge four BISs [see Fig.~\ref{fig6}(a)], of which the innermost spherical-like surface is the static 0-BIS,
and the other three are induced by driving.
Fig.~\ref{fig6}(c) shows the quasienergy spectra with open boundary condition along highly symmetric points $\it\Gamma$-$X_1$-$M$-$\it\Gamma$.
For the 0-gap, there is one surface Dirac cone around the $\it\Gamma$ point and one around $M$, while for the $\pi$-gap, there are two surface Dirac cones around two $X_{1,2}$ points and one at $M$.

The 3D unconventional phase can be characterized by BIS-boundary correspondence
via local $\phi$-structures [see black arrows in $k_z$ direction in Fig.~\ref{fig6}(a)].
Take the quasi-1D system at $k_{x,y}=0$ as an example. The innermost BIS encloses the charge ${\cal O}_1$, forming the topological structure that protects the surface Dirac cone at the $\it\Gamma$ point. Moreover, the spin-momentum locking on the surface [right panel in Fig.~\ref{fig6}(c)] is determined by the SO field of charge ${\cal O}_1$ in the $\gamma_1$-$\gamma_2$ plane [Fig.~\ref{fig6}(b)].
Similarly, the surface modes at $X_1$ ($X_2$, not shown) and $M$ are protected by the 1D topological structures that go through ${\cal O}_2$ $({\cal O}_4)$ and ${\cal O}_3$, respectively~\cite{note_surface}.
In experiment, the detection of 2D surface Dirac cones can also be mapped to the measurement of BIS configurations in the whole Brillouin zone, with which the anomalous chiral phases can be precisely identified.
The 3D anomalous chiral phase in Fig.~\ref{fig6} {\it cannot} be fully characterized by the winding numbers $W_{0,\pi}$. One can find $W_\pi=-1$, but there are three stable surface modes corresponding to valley degree of freedom in the $\pi$-gap (see the details in Appendix~\ref{app4}).
One can expect that the protection from local topological structures will bring about similar surface-geometry dependence and robustness against disorder scattering, similar to the 2D anomalous valley Hall phase analyzed above. These results can be extended to higher dimensions.

Experimentally, the 3D anomalous chiral phase can be simulated using solid-state spin systems such as nitrogen-vacancy center in diamond~\cite{Ji2020},
where the electron and nitrogen nuclear spins can form a two-qubit system coupled by a microwave pulse, mimicking a four-band Dirac model.
In this solid-state spin system, the 3D ${\bm k}$-space Hamiltonian is emulated after mapping $h_i$ ($i=0,1,2,3$)
in Eq.~\eqref{FHam3D} to several experimental parameters controlled by the microwave field  (see, e.g., Ref.~\cite{Ji2020} for details).
Hence, the periodic driving can be generated by engineering the relevant parameters in a particular time-dependent form via a modulated microwave field. 
The new anomalous topology can be detected by spin-resolved quench dynamics, as measured in
Ref.~\cite{Ji2020}, except that a time-periodic Hamiltonian, rather than a static Hamiltonian, is applied after quench.
The experimental simulation of the 3D anomalous chiral phase can also be performed in other solid-state spin systems, such as the nuclear magnetic resonance system~\cite{Xin2020}.

\section{Conclusion and discussion}~\label{sec5}

We have proposed a systematic and highly feasible scheme to realize, manipulate, and detect anomalous Floquet topological phases by engineering BIS configurations.
The scheme works for a class of generic $d$D periodically driven systems and is based on a novel BIS-boundary correspondence shown here.
This correspondence on one hand reveals an equivalence between local topological structure formed in each BIS and gapless modes on the boundary, and maps the detection of ($d-1$)D anomalous boundary modes to the measurement of ($d-1$)D local BIS configurations; on the other hand, opens a systematic way to realize unconventional Floquet topological phases by BIS engineering.
As prime examples, we propose the realization and detection of a 2D anomalous Floquet valley-Hall phase and a 3D anomalous chiral topological phase,
which are both beyond the conventional characterization by global bulk topological invariants, and can be further extended to even higher dimensions. The boundary modes of the unconventional Floquet topological phases exhibit novel edge geometry dependence and are robust against to long-range disorder scatterings.
The proposed models are experimentally well accessible, for their static band structures have been simulated in ultracold atoms~\cite{Wu2016,Sun2017,Liang2021} or solid-state architectures using nitrogen-vacancy center~\cite{Ji2020} and nuclear magnetic resonance~\cite{Xin2020}.

Our theory can be well detected by quench measurements~\cite{Exp2022}.
Take the 2D model as an example.
The different configurations of the ring patterns, including the number of BISs and each BIS enclosing $\it{\Gamma}$ or $M$ point,
can be probed with high-precision by observing the post-quench evolution of a trivial polarized state~\cite{Sun2018,Yi2019}, as also discussed in Sec.~\ref{realization2D}.
The BIS-boundary correspondence enables a full characterization of Floquet topological phases with information of only Floquet bands,
which greatly benefits the quantum simulation of new anomalous Floquet topological bands in ultracold atoms or solid-state spin systems, where the boundary physics is hard to measure.
While we consider here the case that the periodic drive is applied to a static $d$D band structure, the present realization scheme based on BIS engineering also lights up the opportunities of tuning novel Floquet topological phases in other periodically driven systems, which is an interesting direction for our next study.

\section*{Acknowledgements}
We thank Lin Zhang, Jinyi Zhang, Chang-Rui Yi, and Shuai Chen for helpful discussions.
This work was supported by National Key Research and Development Program of China (2021YFA1400900), the National Natural Science Foundation of China (Grants No. 11825401), and the Open Project of Shenzhen Institute of Quantum Science and Engineering (Grant No. SIQSE202003).
L. Z. also acknowledges support from the startup grant of Huazhong University of Science and Technology (Grant No. 3004012191).

\begin{appendix}

\section{Local $\phi$-structures}~\label{app1_new}

The $\phi$-structure is a key concept of our Floquet engineering theory, 
which is responsible for the correspondence between bulk topological structures in the Brillouin
zone and the open boundary energy spectrum.

In our characterization (see Sec.~\ref{sec2} and also Refs.~\cite{Zhanglin2018,Zhanglong2019a}), 
the $d$D bulk topology can be characterized based on the sub-dimensional topology defined on the BISs 
(reflecting the winding of spin-orbit field on the BISs), or equivalently, by the topological charges enclosed by all BISs. 
The BISs are $(d-1)$D momentum subspaces where the band crossing occurs (i.e., $h_0({\bm k})=0$).
The topological charges are dual to the BISs and are located at the nodes of spin-orbit coupling, namely associated with $h_{i}({\bm k})=0$ for $i\neq0$. 
The net topological charge enclosed by a BIS is identical to the $(d-1)$D topology defined on the BIS~\cite{Zhanglong2019a}. In our previous work~\cite{Zhanglin2018,Zhanglong2019a}, it was shown that the topological invariant 
obtained by summing over the contributions from {\it all} BISs determines the global bulk topology of the system.
This bulk-surface duality and related nontrivial topological structures have been experimentally investigated 
by quench dynamics in different simulators~\cite{Ji2020,Xin2020,Yi2019}.


In this work we find that the local topology defined in each single BIS is nontrivial and has unique correspondence to the boundary modes. 
Each fundamental topological structure that a BIS encloses at least a topological charge has its topological consequence.
In particular, under open boundary conditions, the momentum $k_\perp$ perpendicular to the boundary is no longer a good quantum number, but the system can be described as a summation of quasi-1D systems, each with a given ${\bm k}_\parallel$ (the momentum parallel to the boundary). 
A quasi-1D system can host topological edge states (not necessarily of zero energy) if its 1D momentum line ($k_\perp$)
crosses a local BIS-encloses-charge structure. Moreover, when this 1D $k_\perp$ line passes through a topological charge enclosed by the BIS, 
the corresponding edge state has zero energy (see Appendix~\ref{app1}). 
These features characterize the BIS-boundary correspondence and determine the emergence of boundary modes. 
To well capture this result, we define the $\phi$-structure as a topological structure that the 1D $k_\perp$ line crosses a $(d-1)$D BIS 
and meanwhile passes through at least one enclosed charge, as depicted in Fig.~\ref{fig1}. 
This $\phi$-structure must correspond to the zero-energy bound states at both edges, 
and accordingly the $(d-1)$D boundary modes of the original $d$D system, 
with the number and chirality of the boundary modes being determined by the net charge crossed by the $k_\perp$ line (see the proof in the main text).

We would like to point out that while the existence of boundary modes can be predicted by the $\phi$-structures, their robustness, however, requires further protection from 
symmetry or valley degrees of freedom, as demonstrated in Sec.~\ref{zigzag}.

\section{Zero-energy edge states}~\label{app1}

We examine the zero-energy edge states in the 1D subsystem described by the unit vector field ${\bm h}(k_\perp)=(h_0,h_l)=(-\cos k_\perp,\sin k_\perp)$, which
has a $\phi$-structure that one BIS encloses a charge with the reduced value ${\cal C}_\perp=+1$.
Near the charge at $k_\perp^c=0$, the 1D Hamiltonian can be written as
\begin{align}
H_{\rm 1D}(r_\perp)=-\gamma_0\left(1+\frac{1}{2}\frac{\partial^2}{\partial r_\perp^2}+\ui\gamma_0\gamma_l\frac{\partial}{\partial r_\perp}\right),
\end{align}
where $r_\perp$ is the spatial direction perpendicular to the surface.
Suppose $|\Phi_{L,R}(r_\perp)\rangle=\phi_{{L,R}}(r_\perp)|\eta\rangle$ are the two edge states. Here $\phi_{L/R}(r_\perp)$ are the wave functions localized at the left ($L$) or right ($R$) boundary, and $|\eta\rangle$ denotes the spinor.
The eigenvalue equation $H_{\rm 1D}|\Phi_{L/R}\rangle=0$ requires $\ui\gamma_0\gamma_l|\eta\rangle=S|\eta\rangle=\pm|\eta\rangle$ so that
the terms in the parentheses can vanish~\cite{Chiu2013}.

We first consider the left edge state.
Let the left boundary be at $r_\perp=0$ and $\phi_L(r_\perp)$ take the form $e^{-\lambda r_\perp}$.
We then have
$\lambda^2-2\eta\lambda+2=0$,
where $\eta=\pm1$ correspond to $\gamma|\eta\rangle=\pm|\eta\rangle$, respectively. This equation has two solutions $\lambda_\pm=\eta\pm\ui$. Hence,
the unnormalized left edge wave function reads
\begin{align}
|\Phi_{L}(r_\perp)\rangle=\left(c_1 e^{-\ui r_\perp} +c_2 e^{+\ui r_\perp}\right)e^{-\eta r_\perp}|\eta\rangle.
\end{align}
Similarly, one can also obtain the right edge wave function at $r_\perp=L$ ($L$ is system size)
\begin{align}
|\Phi_{R}(r_\perp)\rangle=\left(c_3 e^{-\ui (L-r_\perp)} +c_4 e^{+\ui (L-r_\perp)}\right)e^{\eta (L-r_\perp)}|\eta\rangle.
\end{align}
It is obvious that $\eta$ must be +1 (-1) for the left (right) edge state, i.e., $|\Phi_{L}(r_\perp)\rangle\sim e^{- r_\perp}|+\rangle$ and $|\Phi_{R}(r_\perp)\rangle\sim e^{- (L-r_\perp)}|-\rangle$.

In the same way, one can check the edge states in the system described by ${\bm h}(k_\perp)=(-\cos k_\perp,-\sin k_\perp)$, for which the enclosed charge has ${\cal C}_\perp=-1$.
The 1D Hamiltonian reads
\begin{align}
H_{\rm 1D}(r_\perp)=-\gamma_0\left(1+\frac{1}{2}\frac{\partial^2}{\partial r_\perp^2}-\ui\gamma_o\gamma_l\frac{\partial}{\partial r_\perp}\right),
\end{align}
leading to a reverse result, i.e., $|\Phi_{L}(r_\perp)\rangle\sim e^{- r_\perp}|-\rangle$ and $|\Phi_{L}(r_\perp)\rangle\sim e^{- (L-r_\perp)}|+\rangle$.

\section{Chirality and spin-momentum locking on the boundary}~\label{app2}

The effective Hamiltonian \eqref{Hboundary} for boundary modes is derived from the 1D Hamiltonian $H_{\rm 1D}^\delta(k_\perp)\simeq H_{\rm 1D}(k_\perp)+\sum_{i\neq0,l} h_{i}({\bm k}_\parallel)\gamma_i$. It should be noted that
when there is more than one $\phi$-structure in $k_\perp$ direction, $h_{i\neq 0,l}$ can generally depend on $k_\perp$, making it possible in one $\phi$-structure, we have $h_i({\bm k})\simeq h_i({\bm k}_\parallel)$ and in another, $h_i({\bm k})\simeq -h_i({\bm k}_\parallel)$ for some $i$. In this case, the expression of Eq.~\eqref{Hboundary} no longer holds. Analysis of this complex situation is beyond the scope of this present work. We may discuss it elsewhere.

Here we still consider the case that $\ell\sim\ell_\pm^n$ (each $\phi$-structure encloses a net unit charge with the same sign),
and show how the enclosed charge determines their chirality (for 2D systems) or spin-momentum locked textures (for higher-dimensional systems).
Since ${\bm h}_{\mathrm{so}}({\bm k})$ is linear when approaching a charge at ${\bm k}={\bm k}^c$, we have $h_{i>1}({\bm k})\approx \sum_j\frac{\partial h_i({\bm k}^c)}{\partial k_j}(k_j-k_j^c)$. One can then approximate the dispersion of a boundary mode as a linear Dirac cone centered at ${\bm k}^c$,
with the slope described by a $(d-1)\times(d-1)$ matrix
\begin{align}\label{vmatrix}
v({\bm k}_{\parallel}^c)=
\left(\begin{array}{cccc}
\frac{\partial h_1({\bm k}_\parallel^c)}{\partial k_{\parallel,1}} & \frac{\partial h_1({\bm k}_\parallel^c)}{\partial k_{\parallel,2}}  &\cdots & \frac{\partial h_1({\bm k}_\parallel^c)}{\partial k_{\parallel,d-1}} \\
\frac{\partial h_2({\bm k}_\parallel^c)}{\partial k_{\parallel,1}} & \frac{\partial h_2({\bm k}_\parallel^c)}{\partial k_{\parallel,2}}  &\cdots & \frac{\partial h_2({\bm k}_\parallel^c)}{\partial k_{\parallel,d-1}} \\
\vdots & \vdots &\ddots & \vdots \\
\frac{\partial h_{i\neq l}({\bm k}_\parallel^c)}{\partial k_{\parallel,1}} & \frac{\partial h_{i\neq l}({\bm k}_\parallel^c)}{\partial k_{\parallel,2}}&\cdots &\frac{\partial h_{i\neq l}({\bm k}_\parallel^c)}{\partial k_{\parallel,d-1}}\\
\vdots & \vdots &\ddots & \vdots \\
\frac{\partial h_d({\bm k}_\parallel^c)}{\partial k_{\parallel,1}} & \frac{\partial h_d({\bm k}_\parallel^c)}{\partial k_{\parallel,2}}  &\cdots & \frac{\partial h_d({\bm k}_\parallel^c)}{\partial k_{\parallel,d-1}}
\end{array} \right).
\end{align}
The effective Hamiltonian then takes the form $H_{\rm boundary}^{(\pm)}\simeq \left[v({\bm k}_{\parallel}^c)\delta{\bm k}_\parallel\right]\cdot {\bm\gamma}_P^{(\pm)}$
with $\delta{\bm k}_\parallel={\bm k}_\parallel-{\bm k}_\parallel^c$.
It is obvious that the nonzero elements in Eq.~\eqref{vmatrix} determine how spin and momentum are locked to each other, which are inherited from the Jacobian matrix characterizing the enclosed charge at ${\bm k}^c$ [See Eq.~\eqref{CnJso}].
One can then conclude that spin-momentum locking on the boundary is determined by the total enclosed charge at ${\bm k}_{\parallel}$.

For 2D systems with edges, the bulk Hamiltonian reads $H(k_\perp,k_{\parallel})=h_0(k_\perp)\gamma_0+h_l(k_\perp)\gamma_l+h_S(k_{\parallel})S$ with $S=\ui\gamma_0\gamma_l$.
The SO field reads ${\bm h}_{\rm so}({\bm k})=(h_l(k_\perp),h_S(k_{\parallel}))$ and the charge value at ${\bm k}^c$ is given by
\begin{align}
C={\rm sgn}\left[\frac{\partial(h_l,h_S)}{\partial(k_\perp,k_{\parallel})}\bigg\vert_{{\bm k}={\bm k}^c}\right]={\rm sgn}\left[\frac{\partial h_l(k_\perp^c)}{\partial k_\perp}\cdot \frac{\partial h_{S}(k_\parallel^c)}{\partial k_\parallel}\right].
\end{align}
The effective Hamiltonian for edge modes reduces to $H_{\rm boundary}^{(\pm)}=h_{S}(k_\parallel)P^{(\pm)}SP^{(\pm)}=\pm h_{S}(k_\parallel)P^{(\pm)}$, where we have used $SP^{(\pm)}=\pm P^{(\pm)}$.
According to the analysis in the previous section, the energy of the left edge state is $E_{\rm left}=\pm h_{S}(k_\parallel)={\cal C}_{\perp}h_{S}(k_\parallel)$,
corresponding to the enclosed charge having a reduced value ${\cal C}_{\perp}=\pm1$.
The chirality of edge modes is reflected by their uni-directional propagation direction, which is given by
${\rm sgn}\left(\partial E_{\rm left}/\partial k_\parallel\vert_{k_\parallel=k_\parallel^c}\right)={\rm sgn}\left[{\cal C}_{\perp} \partial h_{S}(k_\parallel^c)/\partial k_\parallel\right]$.
Since ${\cal C}_{\perp}={\rm sgn}\left[\partial h_l(k_\perp^c)/\partial k_\perp\right]$, we have
\begin{align}
{\rm Chirality}={\rm sgn}\left[\frac{\partial h_l(k_\perp^c)}{\partial k_\perp}\cdot \frac{\partial h_{S}(k_\parallel^c)}{\partial k_\parallel}\right]=C.
\end{align}

\section{Derivation of the effective Hamiltonian for subsystems}~\label{app3}

\subsection{Floquet perturbation theory}

We employ the Floquet perturbation theory developed in Refs.~\cite{Mukherjee2020,Haldar2021} to derive the effective Hamiltonian.
Suppose a driven Hamiltonian $H(t)=H_b(t)+H_{\rm so}$, where $H_b(t)$ commutes with itself at different times such
that $H_b(t)|n\rangle=E_n(t)|n\rangle$, with the eigenstates $|n\rangle$ serving as the unperturbed basis,
and $H_{\rm so}$ is a small time-independent perturbation with only off-diagonal entries, i.e., $\langle n|H_{\rm so}|n\rangle=0$ for all $n$.
The ansatz of the Floquet states takes the form
\begin{align}
|{\psi}_n(t)\rangle=\sum_{m} c_m(t)e^{-\ui\int_0^t\ud t' E_m(t')}|m\rangle,
\end{align}
which satisfies
\begin{align}~\label{Floq_c}
|{\psi}_n(T)\rangle=U(T)|{\psi}_n(0)\rangle=e^{-\ui H_FT}|{\psi}_n(0)\rangle.
\end{align}
The time-dependent Schr\"odinger equation $\ui\partial_t|{\psi}_n(t)\rangle=H(t)|{\psi}_n(t)\rangle$ then reduces to
\begin{align}\label{Scheq_c}
\ui \frac{\ud}{\ud t}c_n(t)=\sum_{m\neq n} \langle n|H_{\rm so}|m\rangle e^{\ui\int_0^t\ud t' [E_n(t')-E_m(t')]}c_m(t).
\end{align}

We consider the {\it degenerate} case that the eigenvalues $E_n(t)$ satisfy
\begin{equation}~\label{deg_cond}
e^{\ui\int_0^T\ud t[E_n(t)-E_{n'}(t)]}=1
\end{equation}
for every pair of states.
Up to first order in $H_{\rm so}$, we have the solution to Eq.~\eqref{Scheq_c}
\begin{align}
c_n(T)=c_n(0)-\ui\sum_{m\neq n} \langle n|H_{\rm so}|m\rangle \int_0^T\ud te^{\ui\int_0^t\ud t' [E_n(t')-E_m(t')]}c_m(0),
\end{align}
which gives
\begin{align}~\label{cT_M}
c(T)=(\id-\ui M) c(0),
\end{align}
where $c(t)\equiv[c_1(t),c_2(t),\dots,c_n(t)]^\top$ (the superscript $\top$ denotes the transpose), $\id$
is the identity, and $M$ is a matrix whose elements are given by ($m\neq n$)
\begin{align}
M_{nm}=\langle n|H_{\rm so}|m\rangle \int_0^T\ud te^{\ui\int_0^t\ud t' [E_n(t')-E_m(t')]}
\end{align}

Under the degenerate condition~\eqref{deg_cond}, Eq.~\eqref{Floq_c} leads to
\begin{align}
c(T)=e^{\ui [H_FT-\int_0^T\ud t H_b(t)]}c(0).
\end{align}
Comparing this with Eq.~\eqref{cT_M}, we obtain an expression for the Floquet Hamiltonian
\begin{align}
H_F=\frac{1}{T}\int_0^T\ud t H_b(t)+\frac{M}{T}.
\end{align}

\subsection{The effective Hamiltonian}

For a general Hamiltonian described by Eq~\eqref{FHam0}, we have $H_b(t)=[h_0+V(t)]\gamma_0$ and $H_{\rm so}=\sum_{i>1}h_i\gamma_i$.
On a a driving-induced BIS defined by $h_0({\bm k})=m\omega/2$ with $m>0$, the effective Hamiltonian is given by
\begin{align}
H_{\rm eff}=&\left(\frac{m\omega}{2}+\frac{1}{T}\int_0^T\ud t V(t)\right)\gamma_0+\sum_{n_\uparrow,n_\downarrow}\bigg(\frac{\langle n_\uparrow|H_{\rm so}| n_\downarrow\rangle}{T}\nonumber\\
&\int_0^T\ud te^{\ui\int_0^t\ud t' [m\omega+2V(t')]}|n_\uparrow\rangle\langle n_\downarrow|+{\rm h.c.}\bigg).
\end{align}
Without loss of generality, we assume $V(t)=2\sum_{j=1}^pV^{(j)} \cos(j\omega t)$ with $p$ being a positive integer.
By employing the Jacobi-Anger expansion
\begin{align}
e^{\ui z\sin\theta}=\sum_{n=-\infty}^\infty J_n(z)e^{\ui n\theta},
\end{align}
where $J_n(z)$ is the Bessel function with $J_{-n}(z)=(-1)^nJ_n(z)$,
we then have
\begin{align}\label{Integral_Jn}
\int_0^T\ud te^{\ui\int_0^t\ud t' [m\omega+2V(t')]}&=\int_0^T\ud te^{\ui m\omega t}e^{\ui\sum_{j>0}\frac{4V^{(j)}}{j\omega}\sin(j\omega t)}\nonumber\\
&=\sum_{\sum_{j=1}^pn_j=m}(-1)^m\prod_{j=1}^pJ_{n_j}\left(\frac{4V^{(j)}}{j\omega}\right)T.
\end{align}
Hence, the effective Hamiltonian {\em on} the BIS reads
\begin{align}~\label{Heff_BIS}
\begin{split}
&H_{\rm eff}=\frac{m\omega}{2}\gamma_0+(-1)^m{\cal J}_m\sum_{i>1}h_i\gamma_i, \\
&{\cal J}_m\equiv\sum_{\sum_{j=1}^pn_j=m}\prod_{j=1}^pJ_{n_j}\left(\frac{4V^{(j)}}{j\omega}\right),
\end{split}
\end{align}
with $n_1,n_2,\dots,n_p\in\{0,1,2,\dots,m\}$.
By analytic continuation, we can write the effective Hamiltonian for the corresponding subsystem as
\begin{equation}~\label{Heff_sub_S}
H_{\rm eff}^{(m)}=\left(h_0-\frac{m\omega}{2}\right)\gamma_0+(-1)^m{\cal J}_m\sum_{i>1}h_i\gamma_i.
\end{equation}


For a more general driving $V(t)=2\sum_{j=1}^pV^{(j)} \cos(j\omega t+\varphi_j)$ with phase shifts $\varphi_j$, we have
\begin{align}
&\int_0^T\ud te^{\ui\int_0^t\ud t' [m\omega+2V(t')]}\nonumber\\
=&\int_0^T\ud te^{\ui m\omega t}e^{\ui\sum_{j>0}\frac{4V^{(j)}}{j\omega}[\sin(j\omega t+\varphi_j)-\sin\varphi_j]}\nonumber\\
=&e^{-\ui\sum_{j>0}\frac{4V^{(j)}}{j\omega}\sin\varphi_j}\sum_{\sum_{j=1}^pn_j=m}(-1)^m e^{-\ui\sum_j n_j\varphi_j}\prod_{j=1}^pJ_{n_j}\left(\frac{4V^{(j)}}{j\omega}\right)T.
\end{align}
Compared to Eq.~\eqref{Integral_Jn}, the only difference is that we need to add two phase factors $e^{-\ui\sum_{j>0}\frac{4V^{(j)}}{j\omega}\sin\varphi_j}$ and $e^{-\ui\sum_j n_j\varphi_j}$ to ${\cal J}_m$ in Eq.~\eqref{Heff_BIS}, whose effect is to rotate the SO field, but leave the topology unchanged.

\section{Zigzag edge states of the 2D model}~\label{app_new}

For the zigzag edge geometry illustrated in Fig.~\ref{fig4}(a),
the 2D driven model lives on a square lattice in a diagonal strip.
We define the diagonal directions $k_{1,2}=(k_x\pm k_y)/\sqrt{2}$.
In $k_1$-$k_2$ coordinates,
the static Hamiltonian reads
\begin{align}
H_s(k_1, k_2)&=2\sqrt{2}t_{\rm so}\sin\frac{k_1}{\sqrt{2}}\cos\frac{k_2}{\sqrt{2}}\sigma_X+
2\sqrt{2}t_{\rm so}\cos\frac{k_1}{\sqrt{2}} \nonumber \\
&\sin\frac{k_2}{\sqrt{2}}\sigma_Y+
\left(m_z-4t_0\cos\frac{k_1}{\sqrt{2}}\cos\frac{k_2}{\sqrt{2}}\right)\sigma_z\nonumber\\
&\equiv h_X\sigma_X+h_Y\sigma_Y+h_z\sigma_z,
\end{align}
where we have defined $\sigma_{X,Y}=(\sigma_x\pm\sigma_y)/\sqrt{2}$.
There are four topological charges located at $(k_1,k_2)=(0,0)$, $(0,\sqrt{2}\pi)$, and $(\pm\pi/\sqrt{2},\pi/\sqrt{2})$.
Consider the quasi-1D system at $k_1=0$, whose Hamiltonian reads
\begin{align}
H_{\rm 1D}(k_2)=\left(m_z-4t_0\cos\frac{k_2}{\sqrt{2}}\right)\sigma_z+2\sqrt{2}t_{\rm so}\sin\frac{k_2}{\sqrt{2}}\sigma_Y.
\end{align}

Near the charge at $k_2=0$, the 1D Hamiltonian can be written as
\begin{align}
H_{\rm 1D}(r_\perp)=\sigma_z\left(m_z-4t_0-t_0\frac{\partial^2}{\partial r_\perp^2}-2t_{\rm so}\sigma_{X}\frac{\partial}{\partial r_\perp}\right).
\end{align}
Suppose $|\Phi_{L,R}(r_\perp)\rangle=\phi_{{L,R}}(r_\perp)|\eta\rangle$ are the two zero-energy edge states.
As the discussion in Appendix~\ref{app1}, we have $\sigma_X|\eta\rangle=\pm|\eta\rangle$,
$\phi_L(r_\perp)\sim e^{-\lambda r_\perp}$, and $\phi_R(r_\perp)\sim e^{-\lambda(L-r_\perp)}$, where
\begin{align}
t_0\lambda^2\mp2t_{\rm so}\eta\lambda-m_z+4t_0=0.
\end{align}
Here ``-'' (``+'') is for the left (right) edge state. Finally, we have the results
\begin{align}\label{zigzag_es0}
\begin{split}
|\Phi_{L}(r_\perp)\rangle&=\left(c_1 e^{\ui k_{G}^{+} r_\perp} +c_2 e^{\ui k_{G}^{-} r_\perp}\right)e^{-\frac{t_{\rm so}}{t_0}r_\perp}|+\rangle,\\
|\Phi_{R}(r_\perp)\rangle&=\left(c_3 e^{\ui k_{G}^{+} (L-r_\perp)} +c_4 e^{\ui k_{G}^{-} (L-r_\perp)}\right)e^{-\frac{t_{\rm so}}{t_0}(L-r_\perp)}|-\rangle,
\end{split}
\end{align}
where
\begin{align}
k_{G}^{\pm}=\pm\frac{1}{t_0}\sqrt{t_0(4t_0-m_z)-t_{\rm so}^2}.
\end{align}
Here $a$ denotes the lattice constant.
In particular, for small $t_{\rm so}$, we have
\begin{align}~\label{kg_pm}
k_{G}^{\pm}\approx\pm\sqrt{4-\frac{m_z}{t_0}}.
\end{align}

Near the charge at $k_2=\sqrt{2}\pi$, the 1D Hamiltonian is rewritten as
\begin{align}
H_{\rm 1D}(r_\perp)=&\sigma_z\left[m_z+4t_0-t_0\left(\sqrt{2}\pi+\ui\frac{\partial}{\partial r_\perp}\right)^2\right.\nonumber\\
&\left.-2t_{\rm so}\ui\sigma_{X}\left(\sqrt{2}\pi+\ui\frac{\partial}{\partial r_\perp}\right)\right]
\end{align}
One can check that the zero-energy edge states are given by.
\begin{align}\label{zigzag_esp}
\begin{split}
|\Phi_{L}(r_\perp)\rangle&=\left(c_1 e^{\ui k_{M}^+ r_\perp} +c_2 e^{\ui k_{M}^- r_\perp}\right)e^{-\frac{t_{\rm so}}{t_0}r_\perp}|+\rangle,\\
|\Phi_{R}(r_\perp)\rangle&=\left(c_3 e^{\ui k_{M}^+ (L-r_\perp)} +c_4 e^{\ui k_{M}^- (L-r_\perp)}\right)e^{-\frac{t_{\rm so}}{t_0}(L-r_\perp)}|-\rangle,
\end{split}
\end{align}
where
\begin{align}
k_{M}^{\pm}=\pm\left(\sqrt{2}\pi-\frac{1}{t_0}\sqrt{t_0(4t_0+m_z)-t_{\rm so}^2}\right).
\end{align}
When $t_{\rm so}$ is small, we have
\begin{align}~\label{km_pm}
k_{M}^{\pm}\approx\pm\left(\sqrt{2}\pi-\sqrt{4+\frac{m_z}{t_0}}\right).
\end{align}

For an edge state corresponding to a driving-induced BIS, we have similar results except that $m_z$ should be replaced by the effective Zeeman constant $m_{\rm eff}=m_z-m\omega/2$ ($m=0,\pm1,\dots$).
Here we take the case shown in Fig.~\ref{fig4}(b) as an example.
The edge modes labeled by ``$L_{01}$'' and ``$L_{p1}$'' are protected by the valley at the $G$ point [$(k_1,k_2)=(0,0)$], and ``$L_{02}$'' and ``$L_{p2}$'' are by the valley at $M$ [$(0,\sqrt{2}\pi)$]. For the $L_{01}$ ($L_{p1}$) mode, the corresponding BIS is the one with $m=0$ ($m=1$) and the characteristic momenta are given by Eq.~\eqref{kg_pm}, which yields $k_2^\pm=\pm1$ ($k_2^\pm=\pm\sqrt{3}$).
For the $L_{02}$ ($L_{p2}$) mode, the corresponding BIS is the one with $m=2$ ($m=3$) and the characteristic momenta are given by Eq.~\eqref{km_pm}, which yields $k_2^\pm=\pm(\sqrt{2}\pi-\sqrt{3})$ [$k_2^\pm=\pm(\sqrt{2}\pi-1)$].
These estimated values are close to the numerical results shown in Fig.~\ref{fig4}(d).

\begin{figure*}[t]
\includegraphics[width=0.9\textwidth]{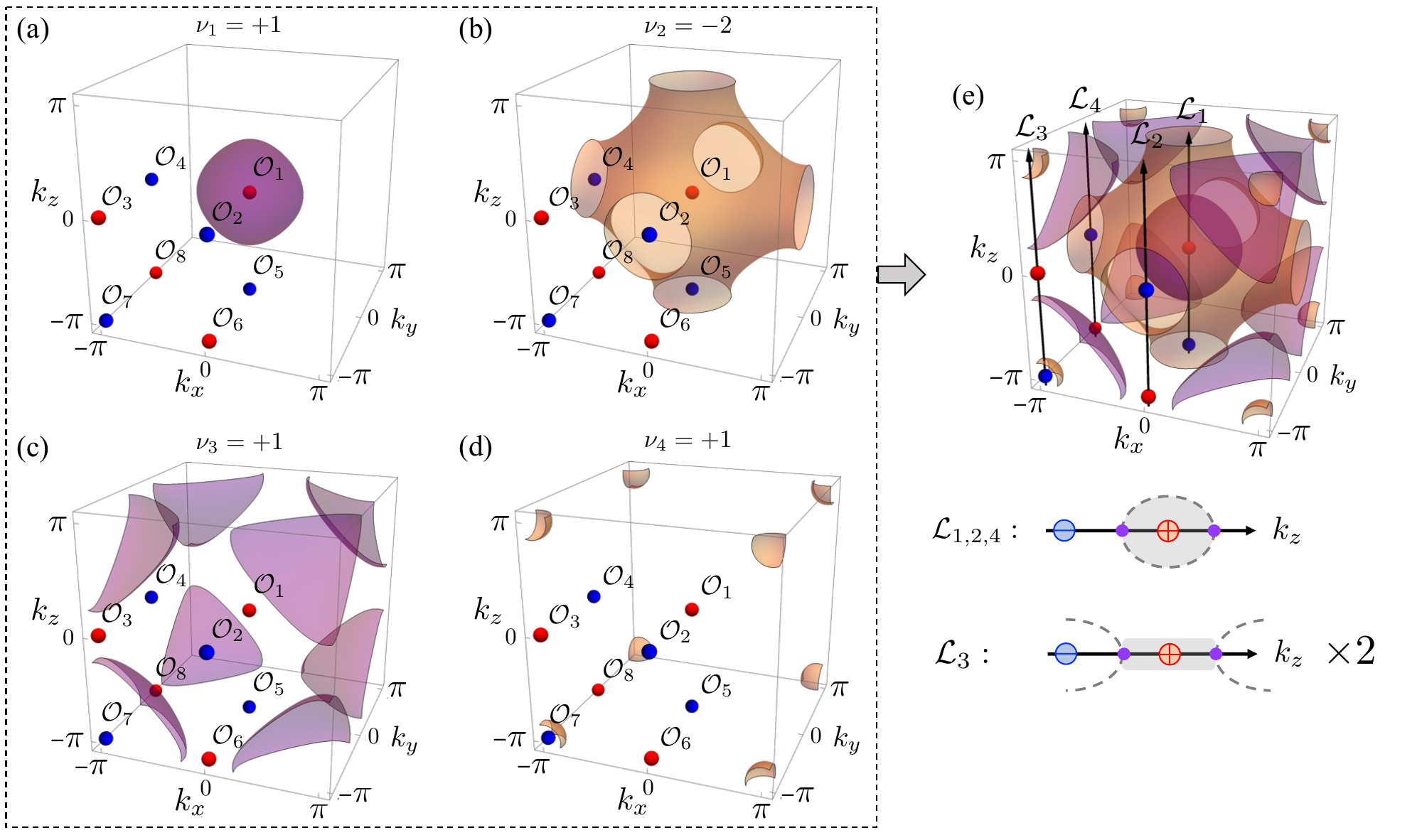}
\caption{BIS-boundary correspondence in the 3D driven model.
(a-d) The BIS characterization for the static topological phases with $m_0=4t_0$ (a), $0.75t_0$ (b), $-2.5t_0$ (c), and $-5.75t_0$ (d).
For each case, the BIS encloses at least one topological charge, with the net charge value defining the topological invariant $\nu_j$ ($j=1,2,3,4$) associated with the BIS.  Here the topological charge with a value ${\cal C}=+1$ ($-1$) is colored in red (blue).
(e) When taking $m_0=4t_0$ and $\omega=6.5t_0$, the Floquet topology is contributed from all static subsystems shown in (a-d).
The BIS configuration forms four nontrivial quasi-1D systems labeled by ${\cal L}_{1,2,3,4}$. For ${\cal L}_{1,2,4}$, each exhibits a $\phi$-structure, while there are two $\phi$-structures in ${\cal L}_{3}$. In the lower panel, the charge in red (blue) has a projective value $C_{\perp}\equiv{\rm sgn}\left(\partial h_3/\partial k_z\vert_{k_z=k_z^c}\right)=+1$ ($-1$). The shaded region denotes the region where $h_0<0$ for each subsystem.
}\label{figS1}
\end{figure*}

\section{BIS-boundary correspondence in the 3D model}~\label{app4}

In this Appendix, we provide more details of the 3D periodically driven model.

We first consider the static Hamiltonian. In the absence of periodic driving, the static topological phases are classified by 3D winding numbers and protected by a chiral symmetry $S=-\gamma_0\gamma_1\gamma_2\gamma_3$. There exist three topological regions:
(I) $2t_0<m_0<6t_0$ with the winding number $W=1$; (II) $-2t_0<m_0<2t_0$ with $W =-2$; and (III) $-6t_0<m_0<-2t_0$ with $W=1$. Using the BIS characterization theory, this 3D model has eight topological charges localized at ${\bm k}^c\equiv(k^c_x,k^c_y,k^c_z)$ with $k^c_{x,y,z}=-\pi$ or $0$, and the phase (I) is characterized by the topological structure that a spherical-like BIS encloses the charge ${\cal O}_1$; the topological invariant associated with the BIS is equal to the enclosed charge value $+1$, characterizing the winding number  $W=1$ [see Fig.~\ref{figS1}(a)]. Similarly, the phase (II) is characterized by a BIS surrounding four charges ${\cal O}_{1,2,4,5}$ [Fig.~\ref{figS1}(b)], and the phase (III) corresponds to the enclosing of all charges but ${\cal O}_7$ [Fig.~\ref{figS1}(c) and (d)].

When the periodic driving $V(t)\gamma_0=2V_0\cos\omega t \gamma_0$ is applied, the time-dependent system has the same chiral symmetry, i.e., $SH({\bf k},t)S^{-1}=-H({\bf k},-t)$ and thus $SH_F({\bf k})S^{-1}=-H_F({\bf k})$~\cite{Zhanglong2020}.
The Floquet system also hosts chiral topological phases, which can be identified by the BIS characterization.
Since ${\bm h}_{\rm so}({\bm k}^c)=0$, the topological charges are immune to the driving
and the only effect of the driving is to induce new BISs.
As demonstrated in the main text, the Floquet system can be treated as a family of subsystems. For our chosen parameters, the Floquet topology of the 3D chiral topological phase is contributed from four static effective subsystems, each corresponding to a static or driving-induced BIS, as shown in Fig.~\ref{figS1}. We define $\nu_j$ ($j=1,2,3,4$) as the topological invariants associated with the BISs. According to the analysis above, we have $\nu_1=\nu_3=\nu_4=+1$ and $\nu_2=-2$. The topological numbers characterizing the two quasienergy gaps are given by $W_0=\nu_1+\nu_3=+2$ and $W_\pi=\nu_2+\nu_4=-1$  [see Eq.~\eqref{W_nu}]. However, numerical calculations show that with an open boundary condition in the $z$ direction, there reside three surface Dirac cones in the $\pi$-gap (see Fig.~4 in the main text). Hence, this 3D chiral phase cannot be simply classified by $W_{0,\pi}$, indicating that  it is an unconventional topological phase.

This anomalous chiral phase can be fully characterized by our BIS-boundary correspondence. As shown in Fig.~\ref{figS1}(e), there are four nontrivial quasi-1D systems (label by ${\cal L}_{1,2,3,4}$) at ${\bm k}^c_{\parallel}\equiv(k^c_x,k^c_y)$ with $k^c_{x,y}=-\pi$ or $0$. Each of them exhibits one (for ${\cal L}_{1,2,4}$) or two (${\cal L}_{3}$) topological $\phi$-structures that the $k_z$ line crosses a BIS at two momentum points and meanwhile goes through an enclosed topological charge. Each $\phi$-structure corresponds to a surface Dirac cone in the corresponding gap. Hence, there should be two surface Dirac cones in the $0$-gap and three Dirac cones in the $\pi$-gap, consistent with the numerical results shown in Fig.~\ref{fig6}(c).

\end{appendix}



\begin{thebibliography}{99}
\bibitem{note} Here $\gamma_l$ can be one of $\gamma_{i}$ matrices ($i>1$) in $H_s$,
or a combination of some of them.
In the latter case, the remaining $\gamma_{i}$ matrices can always be recombined into $d-1$ new ones $\tilde{\gamma}_i$ satisfying
$\{\tilde{\gamma}_i,{\tilde\gamma}_j\}= 2\delta_{ij}$ and $\{\tilde{\gamma}_i, \gamma_l\}=0$.
In the following we drop the tilde notation for simplicity, but assume that such a recombination has been considered if needed.
\bibitem{note_surface} Note that for surface Dirac cones in the $\pi$-gap, the spin-momentum locking is reversed (e.g., compare the two Dirac cones at $M$), which is due to the reversed SO coupling in the effective Hamiltonian for $\pi$-BISs [see Eqs.~\eqref{Heff_sub} and~\eqref{Hboundary_F}].
\end{thebibliography}
\end{document}